\documentclass{emulateapj}
\usepackage{amsmath,natbib}
\usepackage{graphicx}
\usepackage{txfonts}
\usepackage{natbib}
\usepackage{color}

\def\ie{{\it i.e.,}\,}
\def\eg{{\it e.g.,}\,}

\def\la{\hbox{\raise.5ex\hbox{$<$} 
    \kern-1.1em\lower.5ex\hbox{$\sim$}}} 
\def\ga{\hbox{\raise.5ex\hbox{$>$} 
    \kern-1.1em\lower.5ex\hbox{$\sim$}}}

\newcommand{\dgr}{\mbox{$^\circ$}}           
\newcommand{\Msun}{\mbox{M$_\odot$\,}}         
\newcommand{\Lsun}{\mbox{$L_\odot$}}         
\newcommand{\cm}{\mbox{\ cm}}                
\newcommand{\cms}{\mbox{\ cm s${}^{-1}$}}    
\newcommand{\cmss}{\mbox{\ cm s${}^{-2}$}}    

\DeclareMathAlphabet{\mathpzc}{OT1}{pzc}{m}{it}

\shorttitle{A new mixing process operating below shell convection zones}
\shortauthors{Moc\'ak et al.}

\begin{document}

\title{A new stellar mixing process operating below shell convection zones following off-center ignition}
\author{M. Moc\'ak\altaffilmark{1},  Casey A.  Meakin\altaffilmark{2,3}, E. M\"uller \altaffilmark{4} \& L. Siess\altaffilmark{1}}
\altaffiltext{1}{Institut d'Astronomie et d'Astrophysique, Universit\'e Libre de Bruxelles, CP 226, 1050 Brussels, Belgium}
\altaffiltext{2}{Steward Observatory, University of Arizona,Tucson, AZ, 85721 USA}
\altaffiltext{3}{Theoretical Division, Los Alamos National Laboratory, Los Alamos, NM 87545 USA}
\altaffiltext{4}{Max-Planck-Institut f\"ur Astrophysik,Postfach 1312, 85741 Garching, Germany}
\email{mmocak@ulb.ac.be} 
 
\begin{abstract}
\par  During most stages of stellar evolution the nuclear burning of lighter to heavier elements results in a radial  composition profile which is stabilizing against buoyant acceleration, with light material residing above heavier material. However, under some circumstances, such as off-center ignition, the composition profile resulting from nuclear burning can be destabilizing, and characterized by an outwardly increasing mean molecular weight.  The potential for instabilities under these circumstances, and the consequences that they may have on stellar structural evolution, remain largely unexplored. In this paper we study the development and evolution of instabilities associated with unstable composition gradients in regions which are initially stable according to linear Schwarzschild and Ledoux criteria. In particular, we explore the mixing taking place under various conditions with multi-dimensional hydrodynamic convection models based on stellar evolutionary calculations of the core helium flash in a 1.25 \Msun star, the core carbon flash in a 9.3\,\Msun star, and of oxygen shell burning in a star with a mass of 23\,\Msun. The results of our simulations reveal a mixing process associated with regions having outwardly increasing mean molecular weight that reside below convection zones. The mixing is not due to overshooting from the convection zone, nor is it due directly to thermohaline mixing which operates on a timescale several orders of magnitude larger than the simulated flows. Instead, the mixing appears to be due to the presence of a wave field induced in the stable layers residing beneath the convection zone which enhances the mixing rate by many orders of magnitude and allows a thermohaline type mixing process to operate on a dynamical, rather than thermal, timescale. The mixing manifests itself in the form of overdense and cold blob-like structures originating from density fluctuations at the lower boundary of convective shell and "shooting" down into the core. They are enriched with nuclearly processed material, hence leaving behind traces of higher mean molecular weight. In these regions we find that initially smooth composition gradients steepen into stair-step like profiles in which homogeneous, mixed regions are separated by composition jumps. These step like profiles are then seen to evolve by a process of interface migration driven by turbulent entrainment.  We discuss our results in terms of related laboratory phenomena and associated theoretical developments.  We also discuss the degree to which the simulated mixing rates depend on the numerical resolution, and what future steps can be taken to capture the mixing rates accurately. 
\end{abstract}



\section{INTRODUCTION}
\label{sect:intro}

\par The hydrodynamic approach to model certain phases of stellar evolution, like \eg a nuclear core flash, by numerically solving the Euler or Navier-Stokes equations is essentially built upon first principles in physics, and thus (almost) parameter-free. This approach is advantageous compared to (1D) stellar evolutionary calculations when modeling phenomena related to buoyancy or shear driven flow instabilities which are inherently multidimensional in nature. A shortcoming of this approach is that it remains impossible to simulate a star over evolutionary timescales.  Instead, multi-dimensional hydrodynamic calculations provide insight into the nature of mixing processes operating over fluid dynamical timescales which can be used to develop more comprehensive mixing models to be used in evolutionary codes. The classic mixing length theory (MLT) for thermal convection, which remains the standard treatment of convection in stellar evolution calculations, is an example of one such mixing model based on, in this case, a very crude picture of turbulent flow.  A growing body of work presenting simulations of stellar interiors \citep[e.g.][]{dle2006,Herwig2006,Herwig2011,MeakinArnett2007,amy2009,Mocak2008,Mocak2009,Mocak2010} offers the promise of moving beyond crude models like MLT by providing theorists with detailed information about the full non-linear development of fluid instabilities under realistic astrophysical conditions.

\par We build on this work in the present paper by providing a detailed account of a new mixing process that we have discovered in simulations of shell convection.  In particular, we present calculations of both helium and carbon shell convection following off-center ignition.  In both of these evolutionary phases we identify a coupling between the convective shell and the layers which reside below the shell.  This coupling leads to mixing below the burning shell that has potentially significant consequences for the star's structure and evolution.

\par The paper is organized as follows. In Sect.\,\ref{sect:inidat} we described our hydrodynamics code and input data and present some general features of the simulated flows. In Sect.\,\ref{sect:results} we provide a detailed account of the new mixing process that we have identified and discuss its properties in the context of other processes commonly encountered in stellar interiors, including penetration and overshoot, semiconvection, and salt fingering. Contact is made with relevant laboratory, geophysical, and fluid mechanics literature. We summarize our findings in Sect.~\ref{sect:sum}, and indicate future research directions.


\section{Initial data and simulation properties}
\label{sect:inidat}

\begin{figure*}
\includegraphics[width=0.33\hsize]{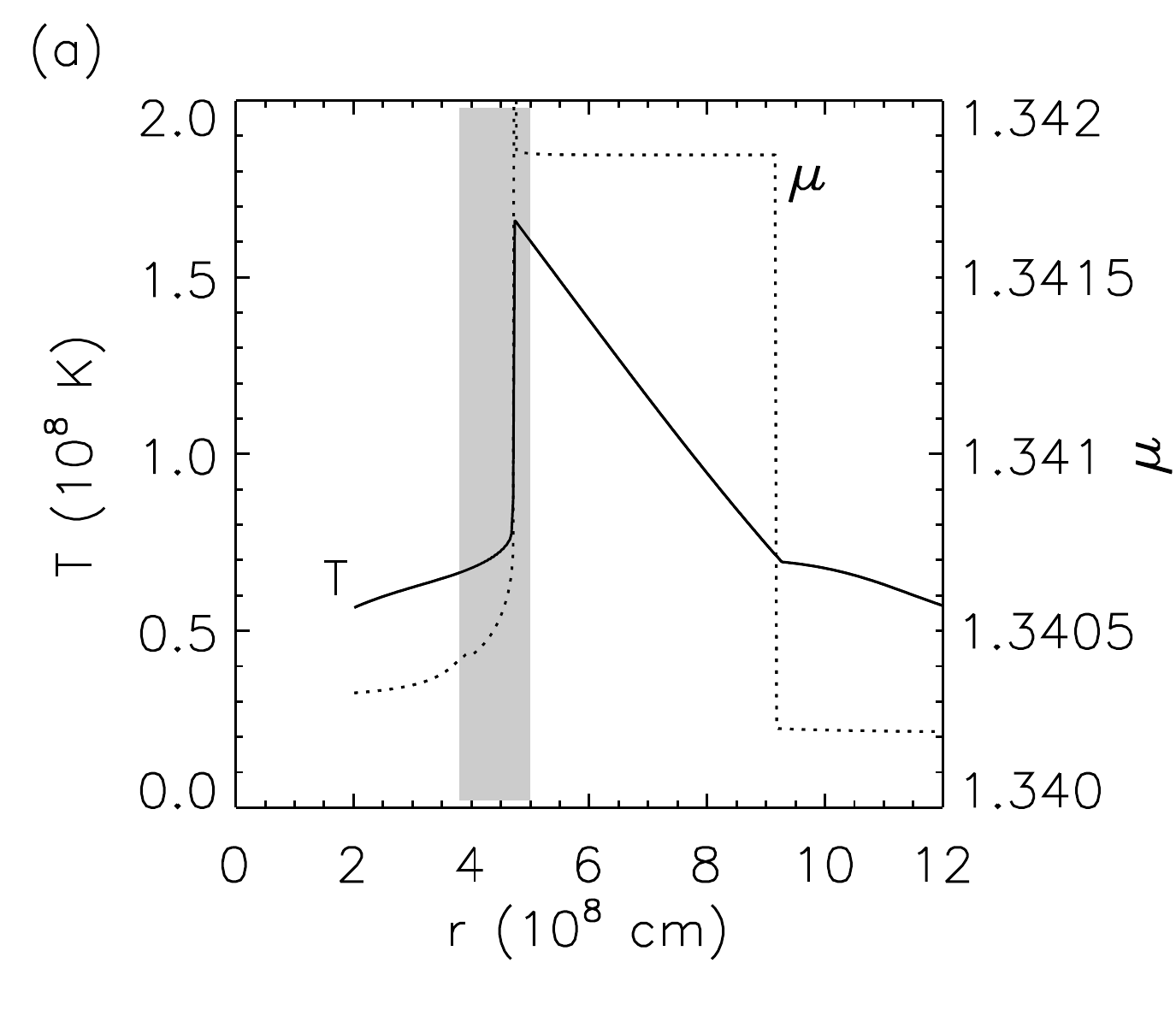}
\includegraphics[width=0.33\hsize]{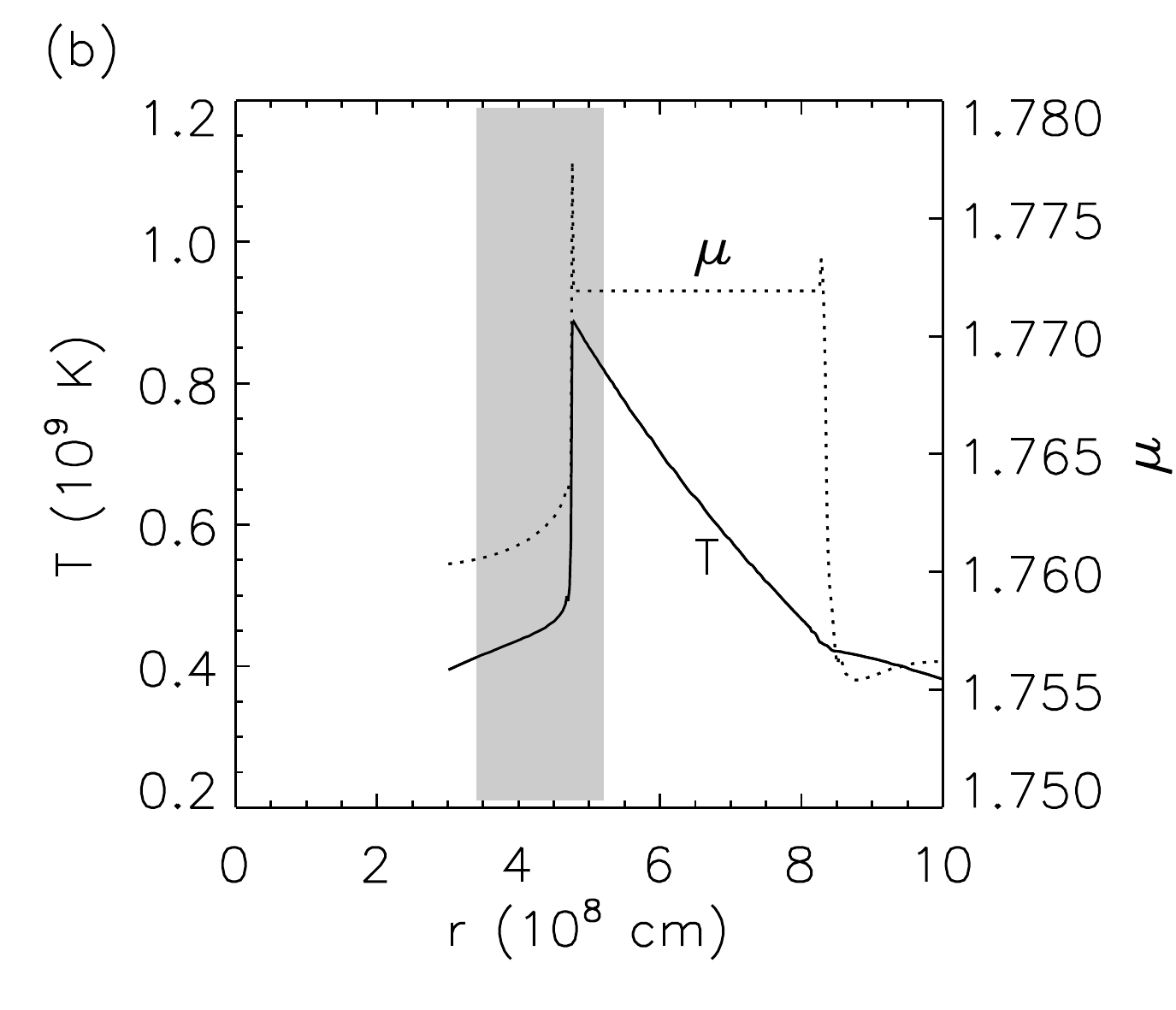}
\includegraphics[width=0.33\hsize]{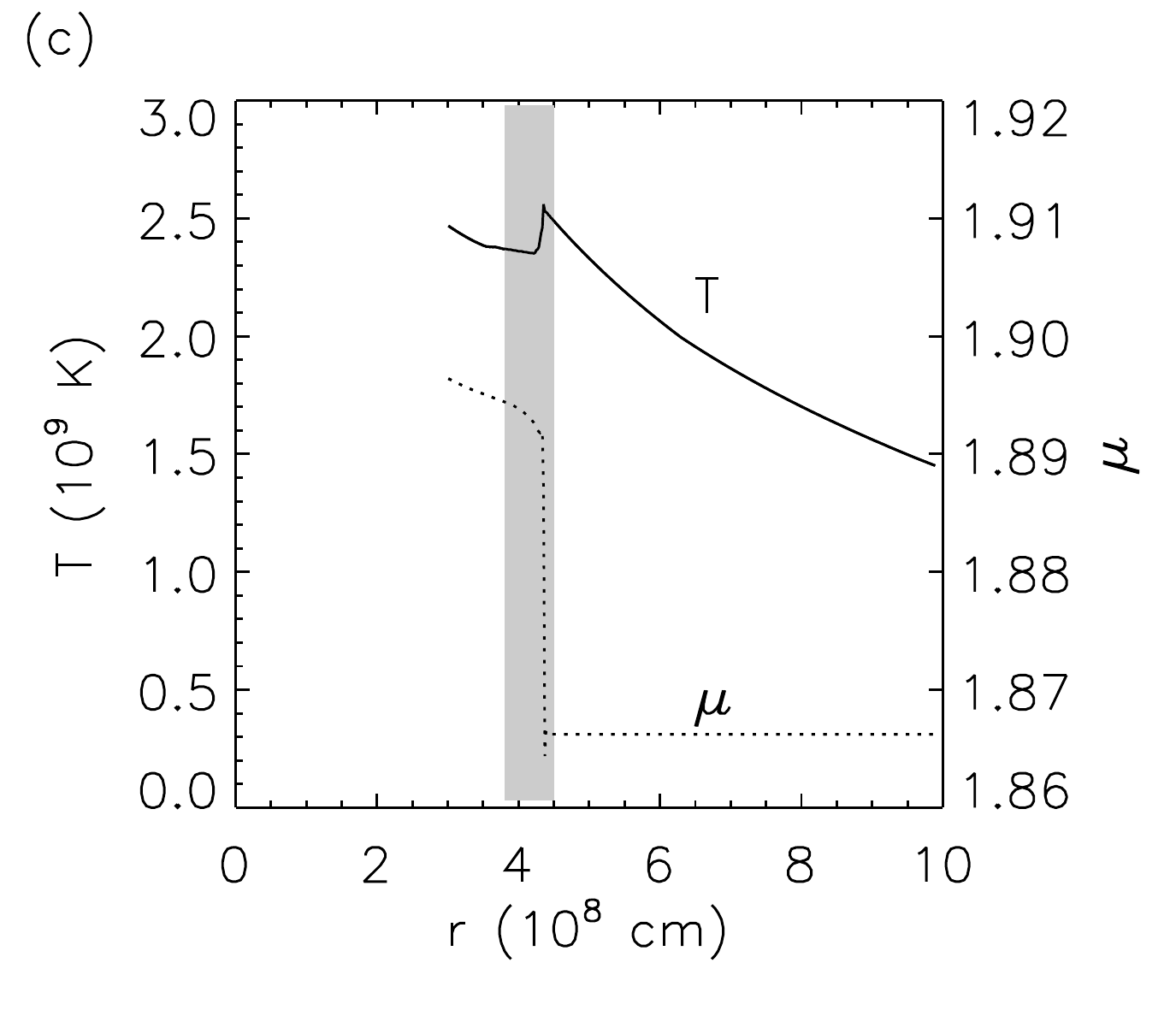}
\caption{Temperature $T$ (solid) and mean molecular weight $\mu$ (dotted) as a function of radius for the initial core helium flash model (a), the core carbon flash model (b), and the oxygen shell burning model (c), respectively . In each of these panels the region of interest below the base of the convection zone is highlighted by the shaded vertical strip. }
\label{fig.inimod}
\end{figure*} 

\begin{figure*} 
\includegraphics[width=0.33\hsize]{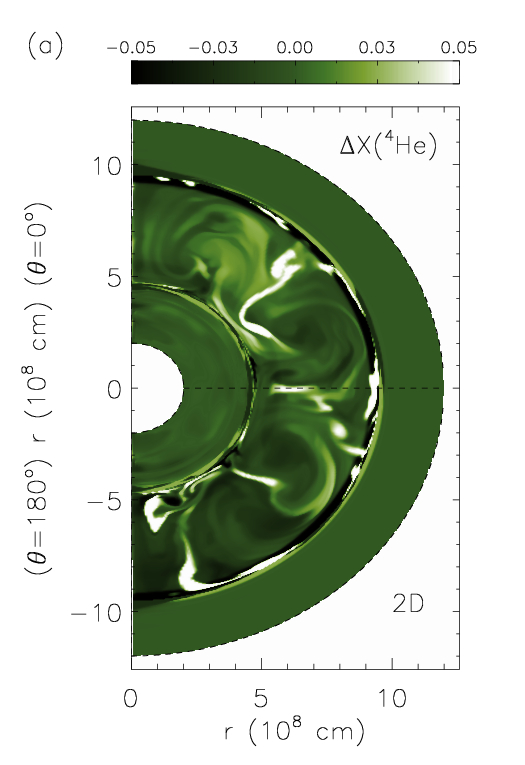}
\includegraphics[width=0.33\hsize]{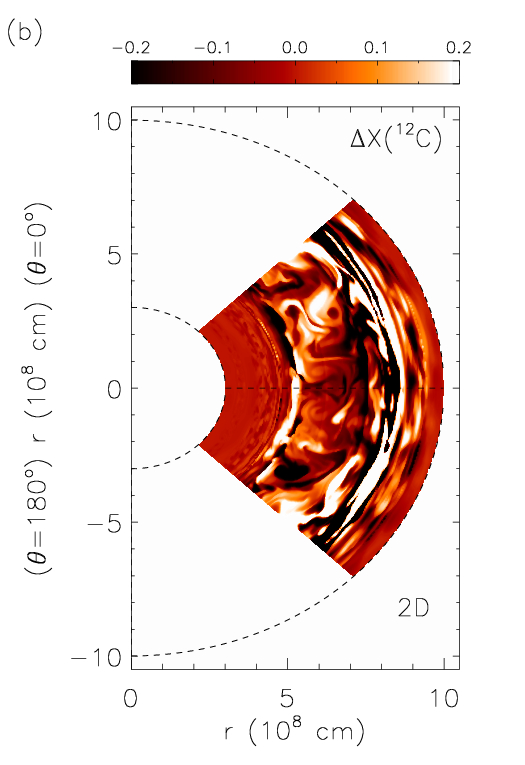}
\includegraphics[width=0.33\hsize]{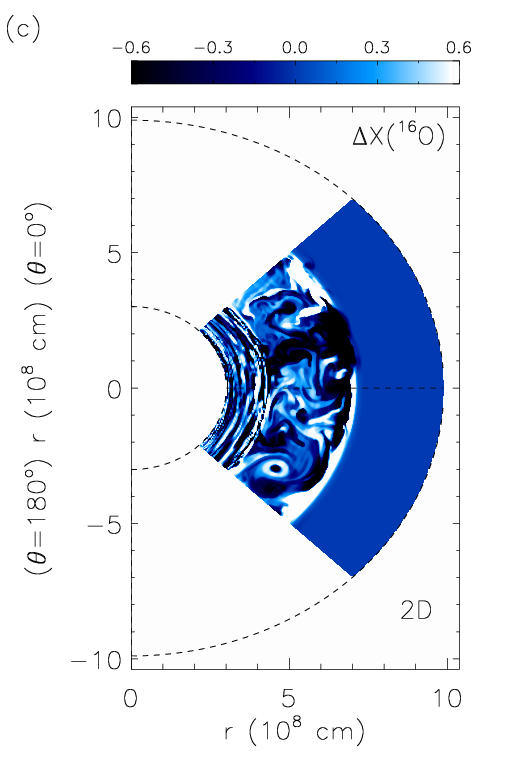}
\caption{Snapshots of the relative angular fluctuations of the mass fraction of $^{4}$He during the core helium flash for model hefl.2d.3 at $t \sim 12000\,$s (a), of $^{12}$C during the core carbon flash for model cafl.2d at $t \sim 682\,$s (b), and of $^{16}$O during oxygen shell burning for model oxfl.2d at $t \sim 940\,$s (c), respectively. The fluctuations are defined by $\Delta X(^{A}{\cal N}) = \mbox{100}\times [X(^{A}{\cal N}) - \langle X(^{A}{\cal N}) \rangle_{\theta}] \,/\, \langle X(^{A}{\cal N}) \rangle_{\theta}$, where $^{A}{\cal N} \in \{ ^4\mbox{He},\,^{12}\mbox{C},\, ^{16}\mbox{O} \}$, and $\langle \rangle_{\theta}$ denotes the angular average at a given radius.}
\label{fig.heflcfloxfl}
\end{figure*} 

\begin{table} 
\caption[]{Some characteristic properties of our initial models: total stellar mass $M$, stellar population, metal content $Z$, mass of the mapped model $M_{m}$, outer (inner) radius $R_{m}$ of the mapped model, and nuclear energy production rate $L_{m}$ of the mapped model, respectively.}
\begin{tabular}{l|lcllll} 
Model & $M$  & Pop. & $Z$ & $M_{m}$  & $R_{m}$    & 
$L_{m}$    \\ 
      & $[\Msun]$ &      &     & $[\Msun]$ & $[10^9\cm]$ &
$[10^9\Lsun]$ \\
\hline 
M  & $1.25$ & I & $~~~0.02$ & $0.45$ & $1.2(0.2)$ & $\sim 1$  \\
L  & $9.3 $ & I & $~~~0.02$ & $0.94$ & $1.0(0.3)$  & $\sim 0.01$ \\
O  & $23. $ & I & $\sim 0.02$ & $2.67$ & $1.0(0.3)$  & $\sim 1000$  \\
\end{tabular} 
\label{imodtab} 
\end{table} 

\subsection{Initial data}

\par We have simulated the reactive hydrodynamic evolution for three phases of stellar evolution.  The initial data used for our calculations include: (1) the helium core of a 1.25 \Msun star during the peak (maximum core luminosity) of the core helium flash computed with the GARSTEC code \citep{WeissSchlattl2000,WeissSchlattl2007}, (2) the carbon-oxygen (C-O) core of a 9.3 \Msun star at peak of the core carbon flash computed with the STAREVOL code \citep{Siess2006}, and (3) the core of a 23 \Msun star during oxygen shell burning \citep{MeakinArnett2007} computed with the TYCHO code \citep{amy2010}.  The thermodynamic structure of all three initial models is qualitatively similar exhibiting an off-center temperature maximum due to nuclear burning in a partially electron degenerate stellar core (see Fig.\,\ref{fig.inimod}).  Some additional properties of the stellar models are included in Table~\ref{imodtab}.

The core helium flash occurs after central hydrogen exhaustion in the semi-degenerate helium core of low-mass stars ($0.7\,\Msun \le M \lesssim 2.2\,\Msun$) due to an off-center ignition of helium by the triple-$\alpha$. The core carbon flash ensues after central helium exhaustion and is also characterized by off-center carbon ignition in a semi-degenerate carbon-oxygen (CO) core of a rather massive star ($7\,\Msun \lesssim M \le 11\,\Msun$), the dominant nuclear reactions being $^{12}$C\,($^{12}$C, $\alpha$)$\,^{20}$Ne and $^{12}$\,C($^{12}$C, p)$\,^{23}$Na followed by $^{16}$O\,($\alpha$, $\gamma$)$\,^{20}$Ne.  Oxygen shell burning, which is typical for massive stars ($M \gtrsim 12\,\Msun$) close to core collapse, is not ignited off-center, but follows an epoch of core oxygen burning which leaves behind a silicon-sulfur rich semi-degenerate core. Because of the steep temperature dependence of the nuclear energy production rates, convection develops in the burning shell.

During the core helium flash the convective shell is enriched mainly in $^{12}$C which results in a negative mean molecular weight gradient below its base (Fig.\,\ref{fig.inimod}\,a)
\footnote{The gradient grows for roughly 10\,000 years since the onset of the core helium flash.}.
The situation is similar during the core carbon flash, where the nuclear ash from carbon burning ($^{20}$Ne and $^{16}$O mainly) increases the mean molecular weight inside its convective shell relative to that of the unburned inner CO core (Fig.\,\ref{fig.inimod}\,b). Oxygen shell burning on the other hand, which follows oxygen core burning, results in a positive composition gradient, \ie the mean molecular weight increases in the direction of gravity at the bottom of the convective shell (Fig.\,\ref{fig.inimod}\,c).

\begin{table} 
\caption{Some properties of the 2D and 3D simulations: number of grid points in $r$ ($N_{r}$), $\theta$ ($N_{\theta}$), and $\phi$ ($N_{\phi}$) direction, radial grid resolution $\Delta r$, angular grid resolution $\Delta \theta$ and $\Delta \phi$ in $theta$ and $phi$-direction, and maximum evolutionary time $t_{max}$ of the simulation, respectively. }
\begin{center}
\begin{tabular}{p{1.1cm}|p{0.4cm}p{0.4cm}p{0.4cm}p{0.8cm}p{0.4cm}p{0.4cm}p{0.8cm}} 
\hline
\hline
model & N$_r$ & N$_\theta$ & N$_\phi$ & $\Delta$r & $\Delta\theta$ & $\Delta\phi$ & t$_{max}$ \\
 & $\#$ & $\#$ & $\#$ & [$10^6$cm] & [$\dgr$] & [$\dgr$] & [10$^3$s]
\\
\hline 
hefl.2d.1 &  180 &   90 &  - & 5.55 & 1.33 &    - &  30 \\
hefl.2d.2 &  270 &  180 &  - & 3.70 & 1    &    - &  30 \\
hefl.2d.3 &  360 &  240 &  - & 2.77 & 0.75 &    - & 120 \\
hefl.3d   &  180 &   90 & 90 & 5.55 & 1.33 & 1.33 &   6 \\
cafl.2d   &  360 &  180 &  - & 1.95 & 0.5  &    - &  60 \\
oxfl.2d   &  400 &  320 &  - & 1.75 & 0.28 &    - &  4.4\\ 
\hline
\end{tabular} 
\end{center}
\label{tab:hydromod}
\end{table}  

\subsection{Properties of hydrodynamic simulations}

The numerical simulations were performed with a modified version of the hydrodynamic code Herakles \citep{Mocak2008}. The code employs the PPM reconstruction scheme \citep{ColellaWoodward1984}, a Riemann solver for real gases according to \citet{ColellaGlaz1984}, and the consistent multi-fluid advection scheme of \citet{PlewaMueller1999}. Self-gravity, thermal transport and nuclear burning are included in the code.  Nuclear reaction networks are generated using the REACLIB library developed by Thielemann (private communication). In Table\,\ref{tab:hydromod} we summarize some characteristic parameters of our 2D and 3D hydrodynamic simulations.

\par The core helium flash simulations were performed on an equidistant spherical polar grid in 2D ($r, \theta$) and 3D ($r, \theta, \phi$). While the angular grid of the axisymmetric 2D simulations hefl.2d.2 and hefl.2d.3 covered the full angular range, \ie $0\dgr \le \theta \le 180\dgr$, the 2D simulation hefl.2d.1 was restricted to the angular region $30\dgr \le \theta \le 150\dgr$.  The 3D simulation hefl.3d was performed within an angular wedge given by $30\dgr \le \theta \le 150\dgr$ and $-60\dgr \le \phi \le +60\dgr$, respectively. This allowed us to simulate the 3D model with a reasonable amount of computational time using a radial and angular resolution comparable to that of the 2D model hefl.2d.1 for several convective turnover timescales. In all core helium flash simulations the radial grid ranged from $r = 2\times 10^8\,$cm to $r = 1.2\times 10^9\,$cm. Boundary conditions were reflective in all directions except for simulations hefl.2d.1 and hefl.3d, where we utilized periodic boundary conditions in angular direction(s) to avoid a numerical bias in case of wide angular convective structures. Abundance changes due to nuclear burning during the He-flash are described by a reaction network consisting of the four $\alpha$-nuclei $^4$He, $^{12}$C, $^{16}$O, and $^{20}$Ne coupled by seven reactions.

The core carbon flash simulation cafl.2d was performed on a 2D equidistant spherical polar grid ($r, \theta$) covering a 90$\dgr$ angular wedge centered at the equator ($\theta = 90\dgr$), and a radial grid ranging from $r = 3\times 10^8\,$cm to $r = 1\times 10^9\,$cm.  Boundary conditions were reflective in radial direction and periodic in angular direction. Abundance changes due to nuclear burning were described by a reaction network consisting of $^1$H, $^4$He, $^{12}$C, $^{14}$N, $^{16}$O,$^{20}$Ne, $^{22}$Ne, $^{23}$Na, and $^{24}$Mg coupled by 17 reactions.

The oxygen shell burning simulation oxfl.2d was performed on a 2D equidistant spherical polar grid covering a 90$\dgr$ wedge centered at the equator, and a radial grid ranging from $r = 2\times 10^8\,$cm to $r = 1\times 10^9\,$cm.  Boundary conditions were reflective in radial direction and periodic in angular direction. Abundance changes due to nuclear burning were described by a reaction network consisting of neutrons, $^1$H, $^4$He, $^{12}$C, $^{16}$O, $^{20}$Ne, $^{23}$Na, $^{24}$Mg, $^{28}$Si, $^{31}$P, $^{32}$S, $^{34}$S, $^{35}$Cl, $^{36}$Ar, $^{38}$Ar, $^{39}$K, $^{40}$Ca, $^{42}$Ca, $^{44}$Ti, $^{46}$Ti, $^{48}$Cr, $^{50}$Cr, $^{52}$Fe, $^{54}$Fe, and $^{56}$Ni coupled by 76 reactions.

The global character of the flow is illustrated in Figure~\ref{fig.heflcfloxfl} which shows a snapshot of the composition fluctuations in the developed convective flows from the hefl.2d.3 (helium burning), cafld.2d (carbon burning), and oxfl.2d (oxygen burning) calculations.  The basic topology of the flow is that of a turbulent convective shell surrounded by stably stratified layers which host internal waves.   In these 2D simulations the shell convection consists of interacting plumes and convective rolls which span the depth of the shell.  In 3D the flow is better described by plumes than rolls, since vortices are no longer pinned to meridional planes by the geometry of the calculation \citep[see e.g.][]{MeakinArnett2007}.  In both the 2D and the 3D simulations, the internal waves can be seen as narrow (in radius) bands extending in angle above and below the turbulent convective shell. The lack of a strong signature of internal waves above the convective shell in the helium and oxygen burning models in Figure~\ref{fig.heflcfloxfl} is a result of the composition profile lacking a gradient there.  The signature of waves is present in other fields which have gradients, such as the pressure, density, and velocity fields.

\begin{figure*}
\includegraphics[width=0.33\hsize]{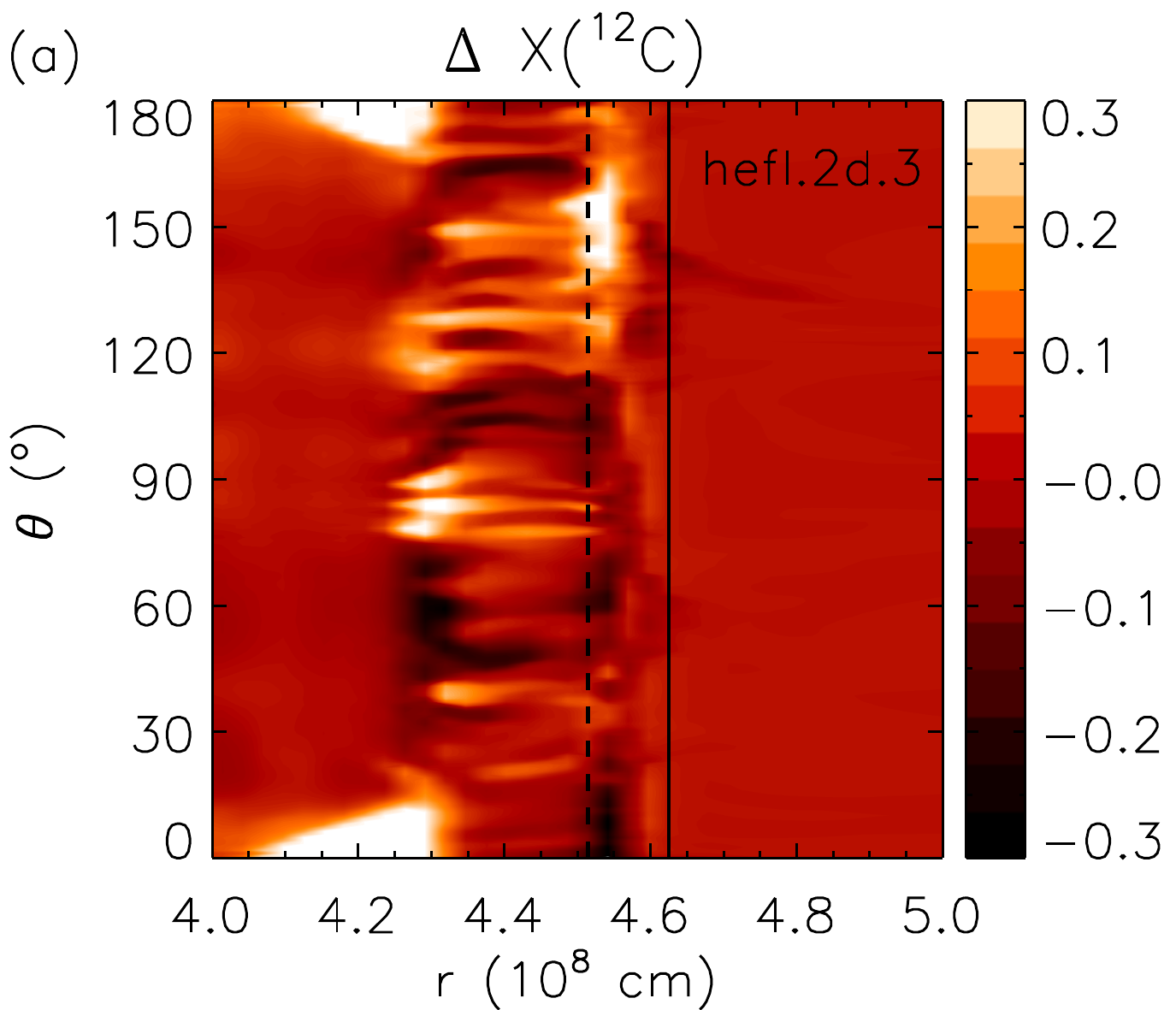}
\includegraphics[width=0.33\hsize]{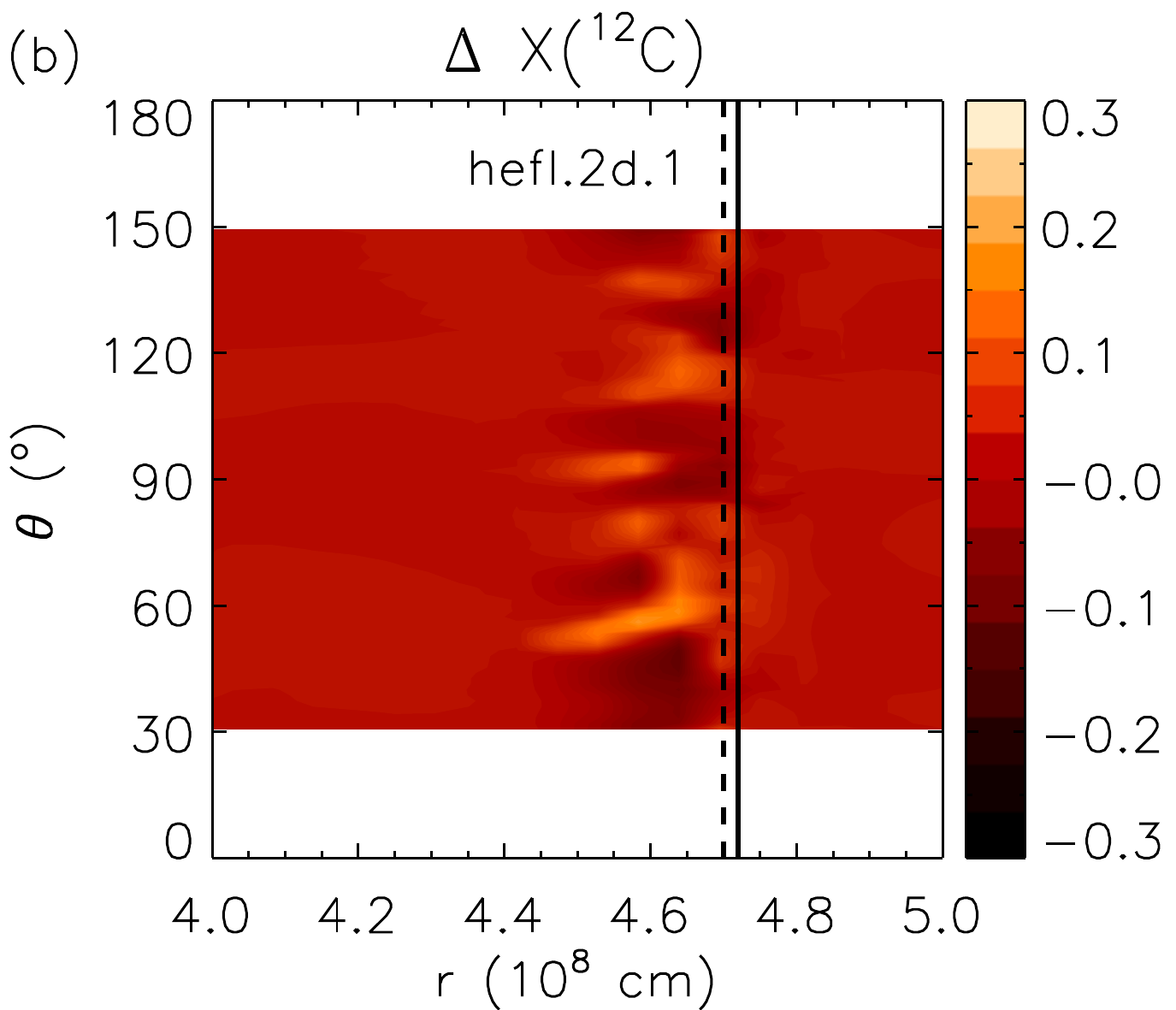}
\includegraphics[width=0.33\hsize]{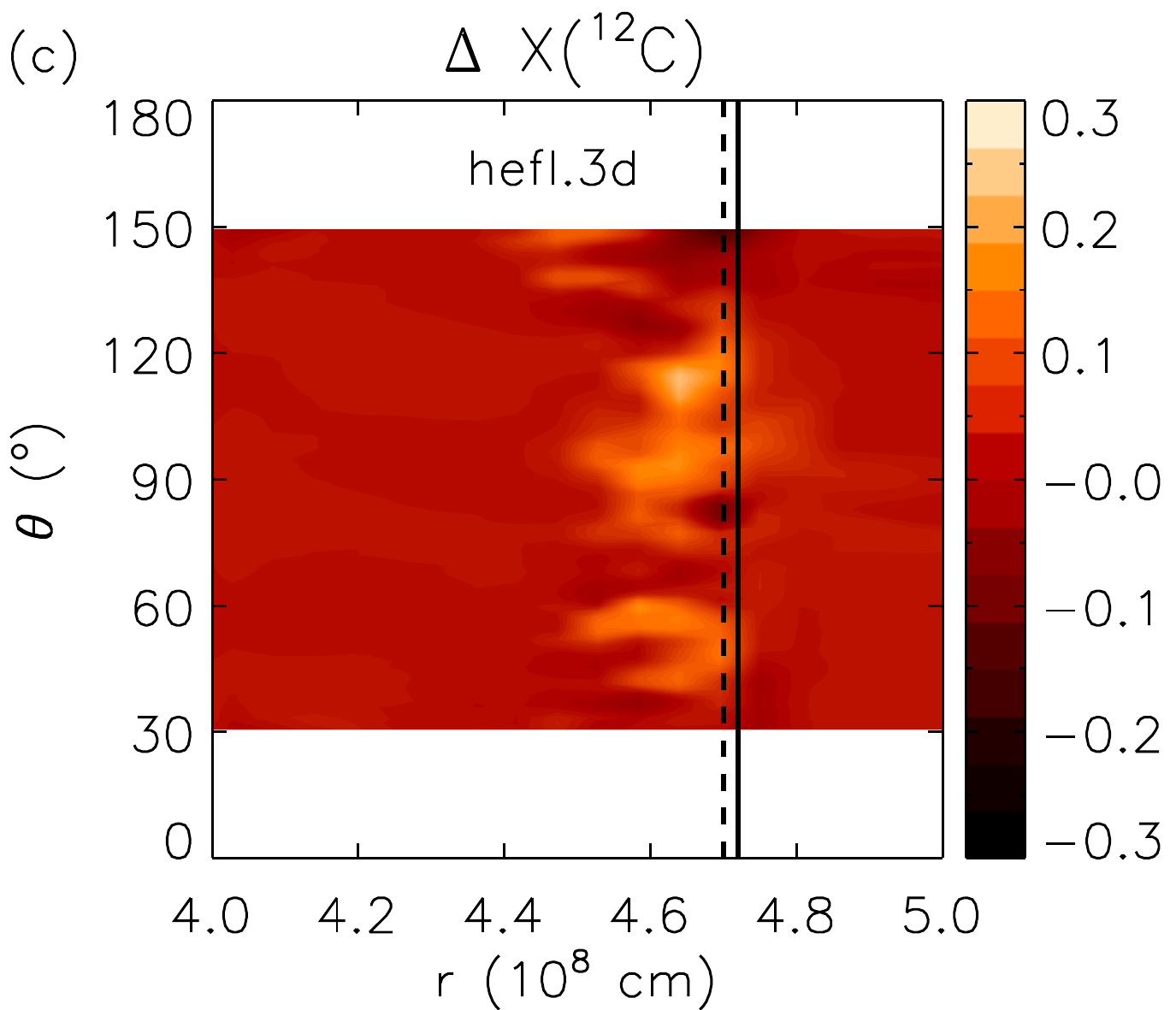}
\caption{Maps of the relative angular fluctuation in carbon mass fraction $\Delta X(^{12}\mbox{C}) \equiv 100\times (X(^{12}\mbox{C}) - \langle X(^{12}\mbox{C}) \rangle_{\theta}) \,/\, \langle X(^{12}\mbox{C})\rangle_{\theta}$ at the bottom of the He-flash convection zone, where $\langle \rangle_{\theta}$ denotes the horizontal average at a given radius.  From left to right, the following models are shown: (left) hefl.2d.3 at $t = 63430\,$s, (middle) hefl.2d.1 at $t = 6000\,$s , and (right) a meridional plane of the 3D model hefl.3d at $t = 6000\,$s. The vertical solid line marks the bottom boundary of the convection zone which is equal to the position of $T_{max}$, and the dashed line gives the location from where our mixing begins. 
}
\label{fig.fingcarb}  
\end{figure*} 

{\it Helium Flash Convection.---} The hydrodynamic properties of shell convection during the core helium flash are described in detail in \citet{Mocak2008, Mocak2009}. Convection starts early ($t < 1000\,$s) and quickly extends over the whole convectively unstable region. In axisymmetry (i.e., 2D), the convection is characterized by fast and large circular vortices, while in 3D the convective flow is less ordered showing slower and smaller turbulent features. The convective velocities in our 3D model match those predicted by MLT quite well ($|v| \sim v_{MLT} \lesssim 1 \times 10^{6} \cms$), whereas the velocities in our 2D models exceed those by up to a factor of four. Turbulent entrainment leads to a growth of the width of the convection zone on a dynamic timescale in both the 2D and 3D models due to an exchange between the potential energy contained in the stratified layers at the boundaries of the convection zone and the kinetic energy of the turbulent flow inside the convection zone.

{\it Carbon Flash Convection. ---} An analysis of the hydrodynamic properties of shell convection during the core carbon flash, based on our 2D model cafl.2d, shows that the convective flow is dominated by small circular vortices.  The angular averaged amplitudes of velocities inside the convection zone $|v| \sim 4\times 10^6 \cms $ exceed those predicted by MLT $v_{MLT} \sim 1.5 \times 10^5 \cms $ by about an order of magnitude. From authors experience, corresponding 3D convection (not calculated) would be characterized by smaller velocities by a factor from 2 to 5. The operation of turbulent entrainment at the convective boundaries has also been found but not investigated in detail in this paper. 

{\it Oxygen Shell Burning. ---} The hydrodynamic properties of shell convection during oxygen shell burning are discussed in detail in \citet{MeakinArnett2007}. Our 2D model oxfl.2d reproduces approximatelly the results of these authors. The angular averaged amplitudes of the convective velocities inferred from model oxfl.2d are $|v| \sim 1\times 10^{7}\cms$, which is roughly equivalent to the velocity $v_{MLT}$ predicted by MLT. Turbulent entrainment has been detected in model oxfld.2d, too, but we have not further analyzed this phenomenon; for more details see \citet{MeakinArnett2007}. 

\section{Results}
\label{sect:results}

\subsection{General Features of the Mixing}
\label{sec:general-features}

\par  In this paper we focus our analysis on a mixing process which operates within the stably stratified layers residing beneath the shell convection zones in both the carbon- and helium-flash simulations.  This mixing process is not observed to operate below the shell convection zone in the oxygen burning case.  In this subsection we describe the salient features of the mixing for the two shell flash calculations, and contrast these with the oxygen shell burning case.

\begin{figure*} 
\includegraphics[width=0.33\hsize]{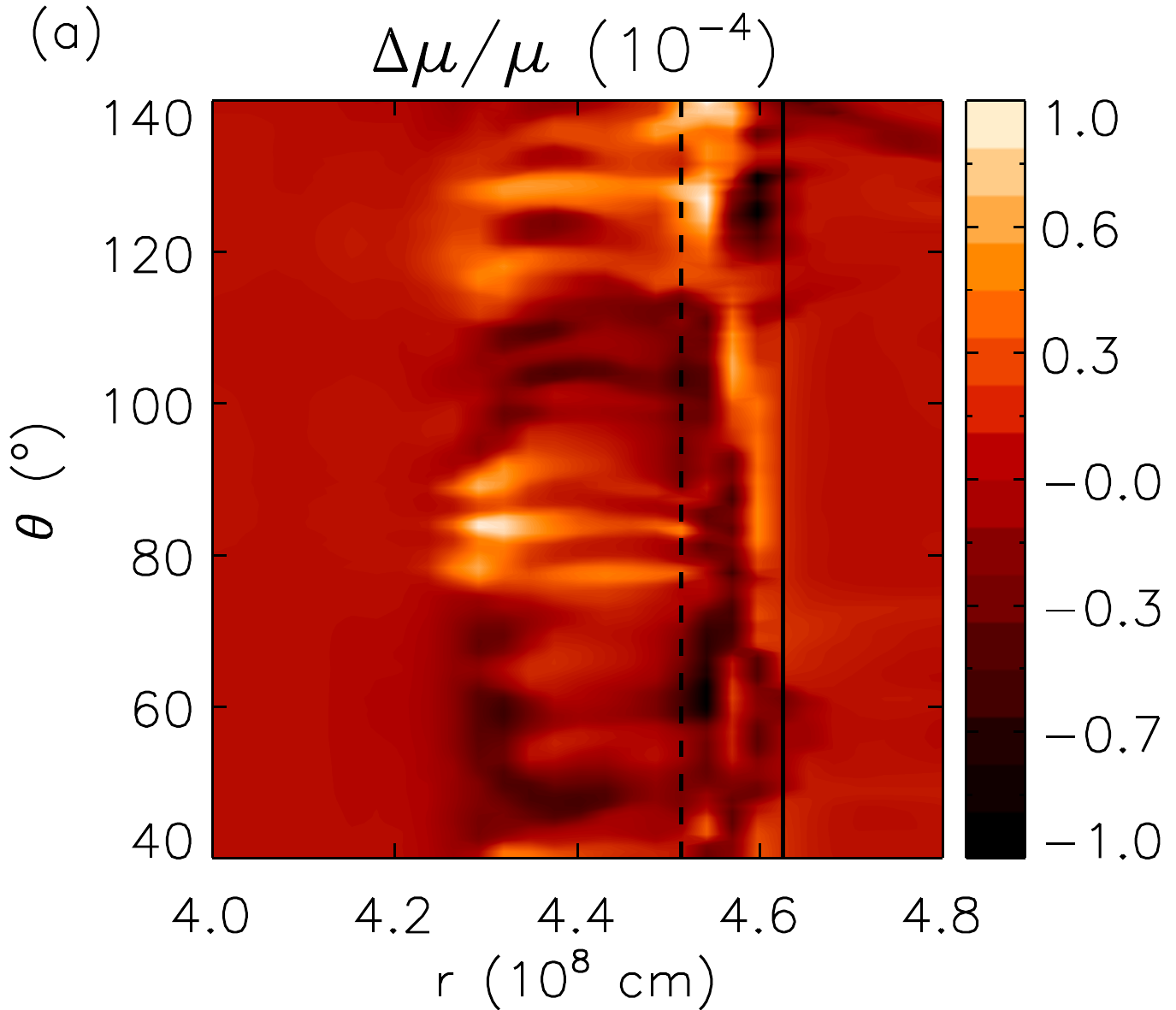}
\includegraphics[width=0.33\hsize]{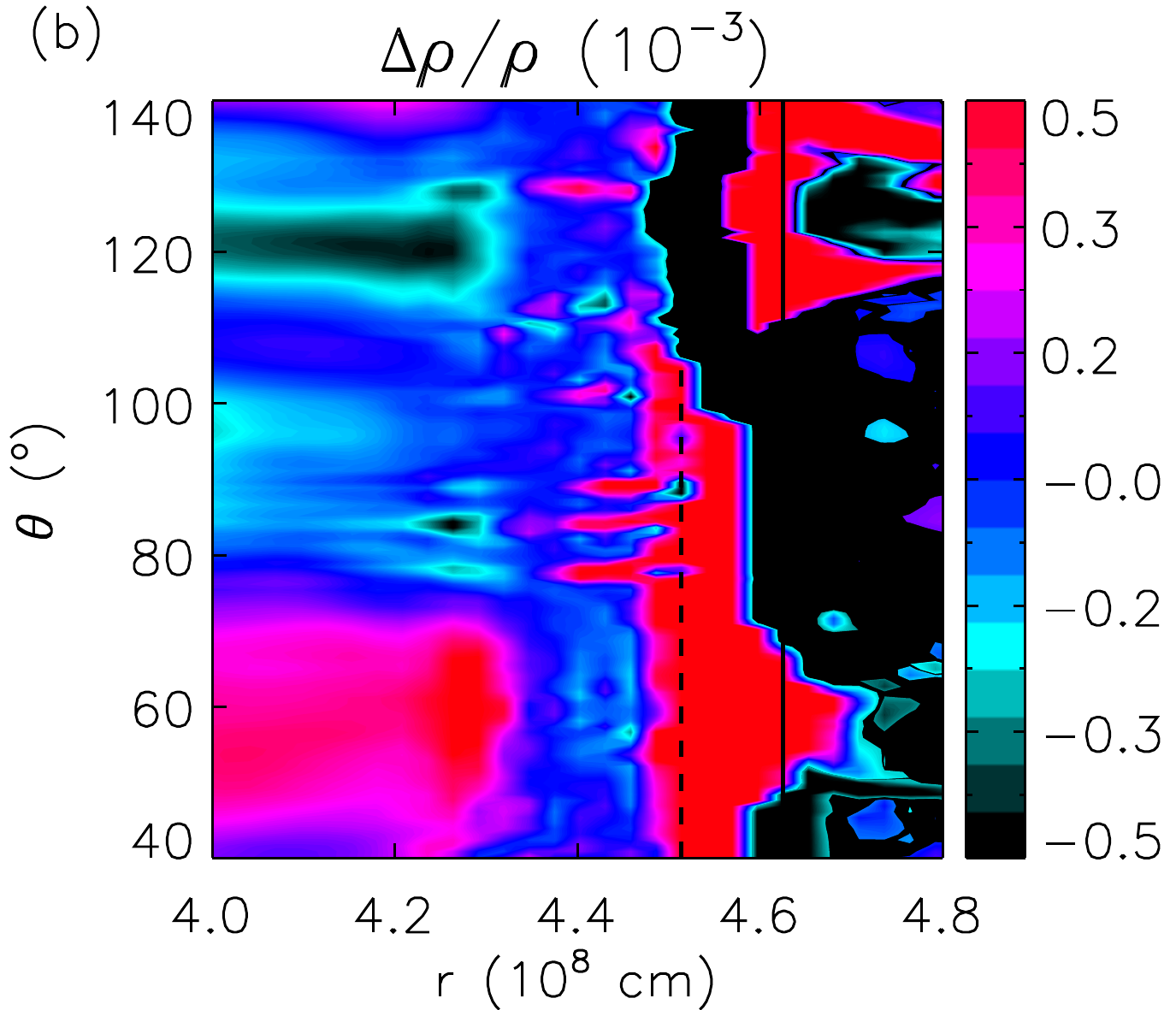}
\includegraphics[width=0.33\hsize]{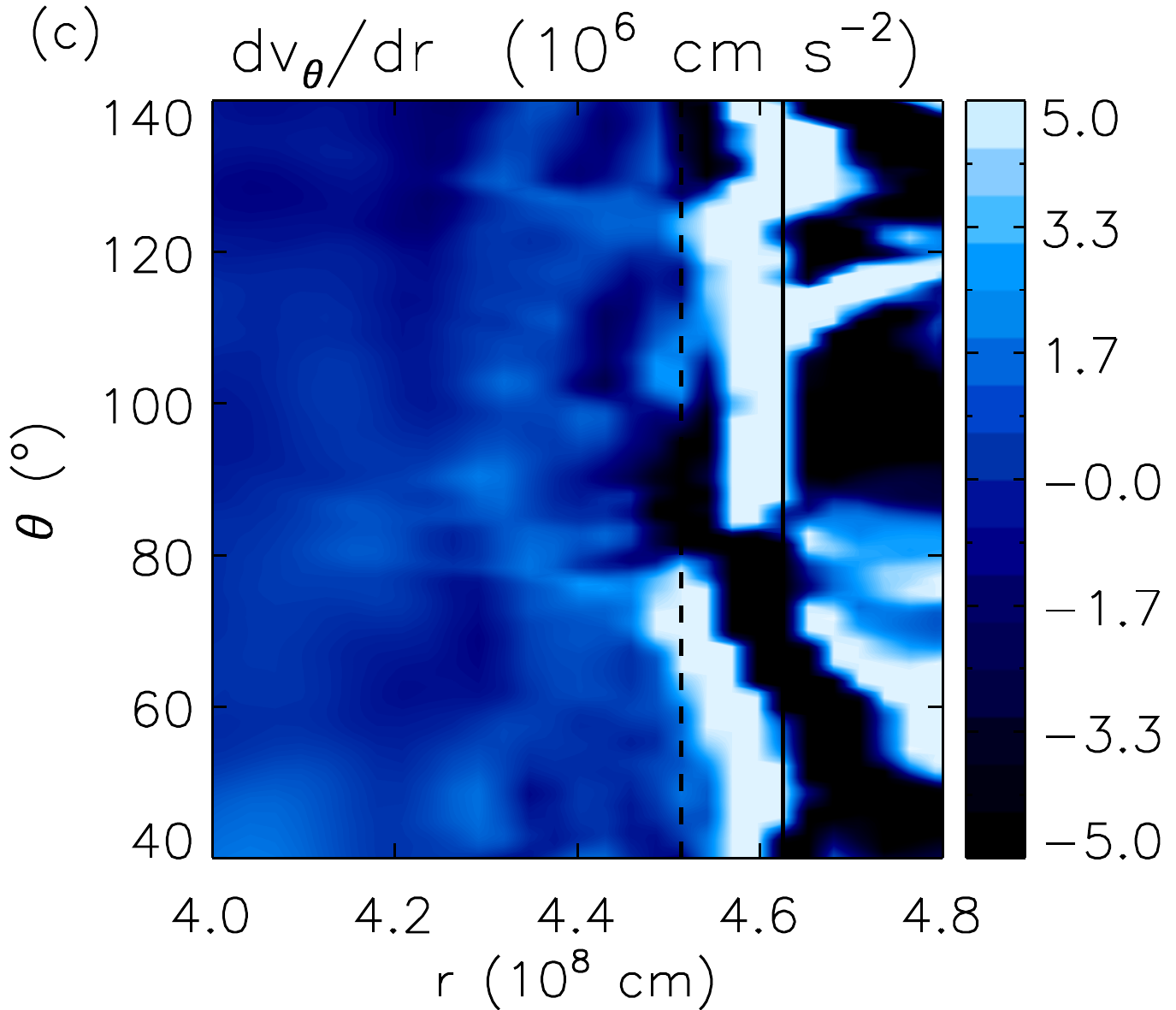} \\
\includegraphics[width=0.33\hsize]{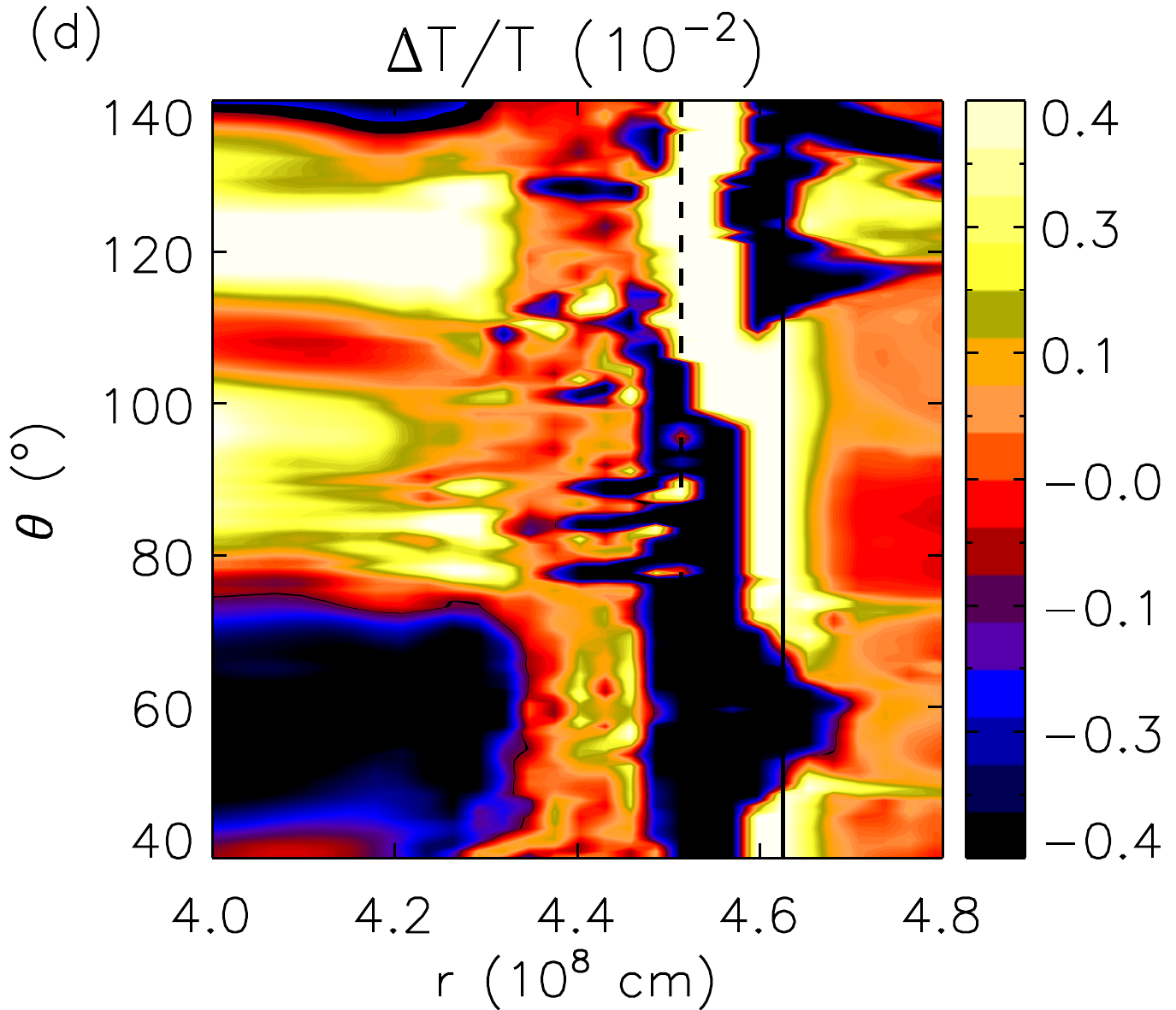}
\includegraphics[width=0.33\hsize]{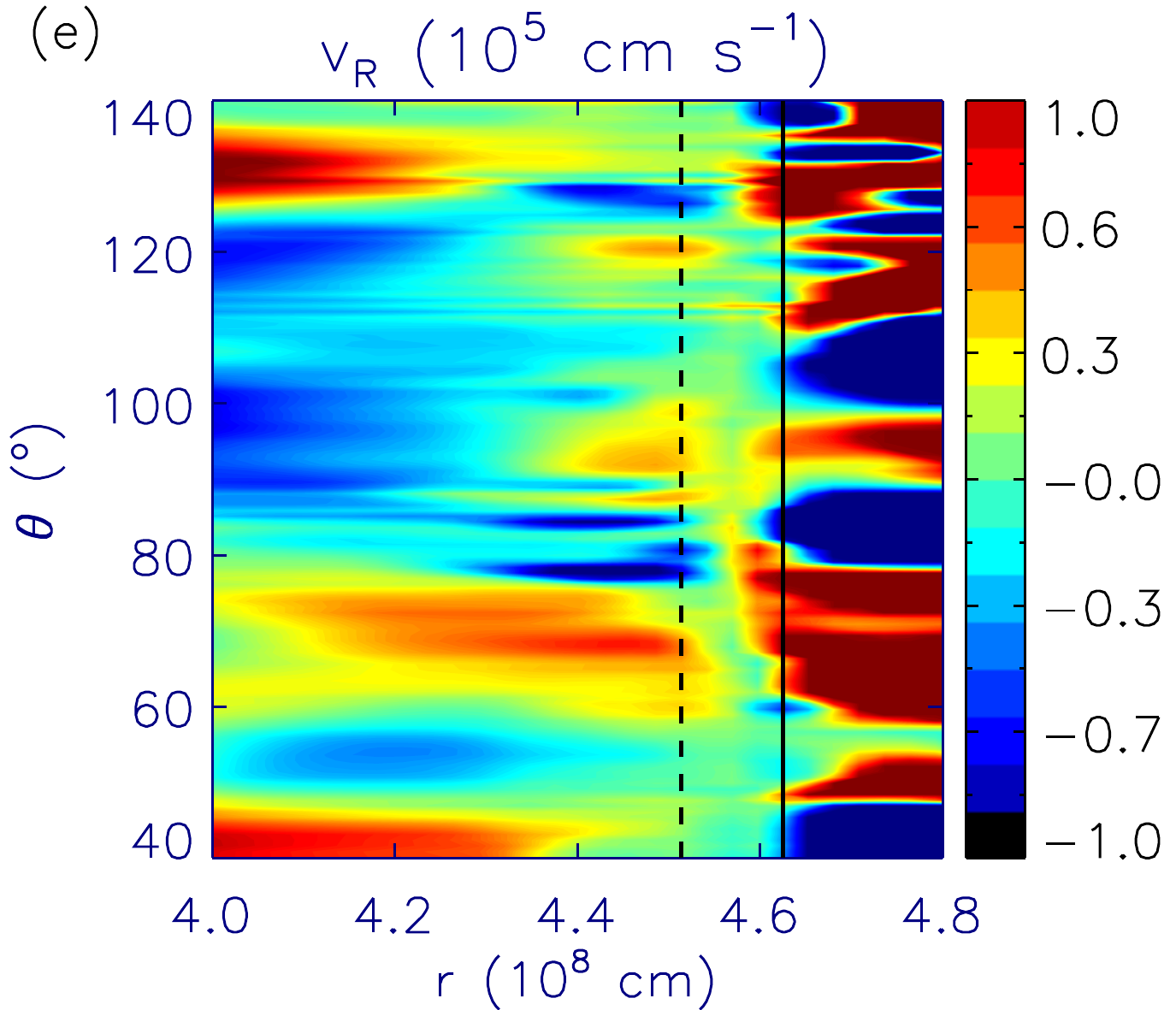}
\includegraphics[width=0.33\hsize]{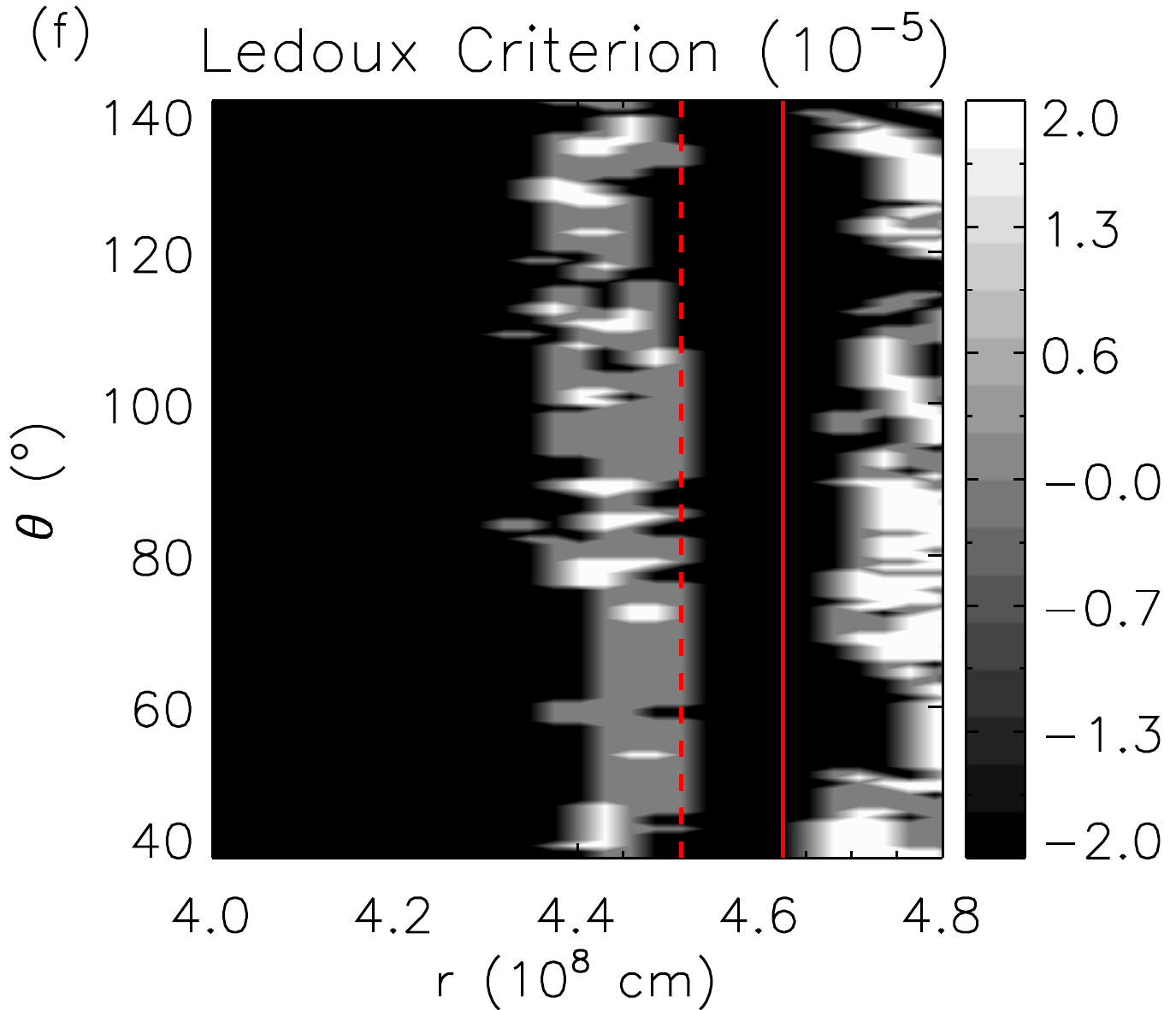}
\caption{Snapshots showing the base of the convection zone in the 2D model hefl.2d.3 at $t = 63430\,$s. The panels (a,b,d) give the relative difference between the local and the horizontally averaged value at a given radius of the mean molecular weight $\mu$, the density $\rho$ and the temperature $T$, respectively. The panels (c,e,f) display the deviation of angular velocity with respect to radius r $dv_\theta/dr$, radial velocity $v_R$, and the Ledoux criterion (a positive value implies that the flow is Ledoux or dynamically unstable; see Eq.\,\ref{eq:stability}), respectively. The vertical solid line marks the bottom of the convection zone (location of $T_{max}$), while the dashed line gives the location from where our mixing begins. Note that only a part of the computational domain is shown. }
\label{fig.fingall}  
\end{figure*} 

\par {\it Helium Flash Convection.---} In the helium-flash convection simulations we find a mixing process taking place below the temperature maximum immediately after convection starts.  The mixing manifests as the formation of blob-like structures in the stable layer just below the temperature maximum, which fall inward toward the stellar center.  These structures are present in both the 2D and 3D simulations and can be seen in the snapshots presented in Figure~\ref{fig.fingcarb} which shows the fluctuations in the helium burning product C$^{12}$ for three of our He-flash simulations.  

\begin{figure*} 
\includegraphics[width=0.325\hsize]{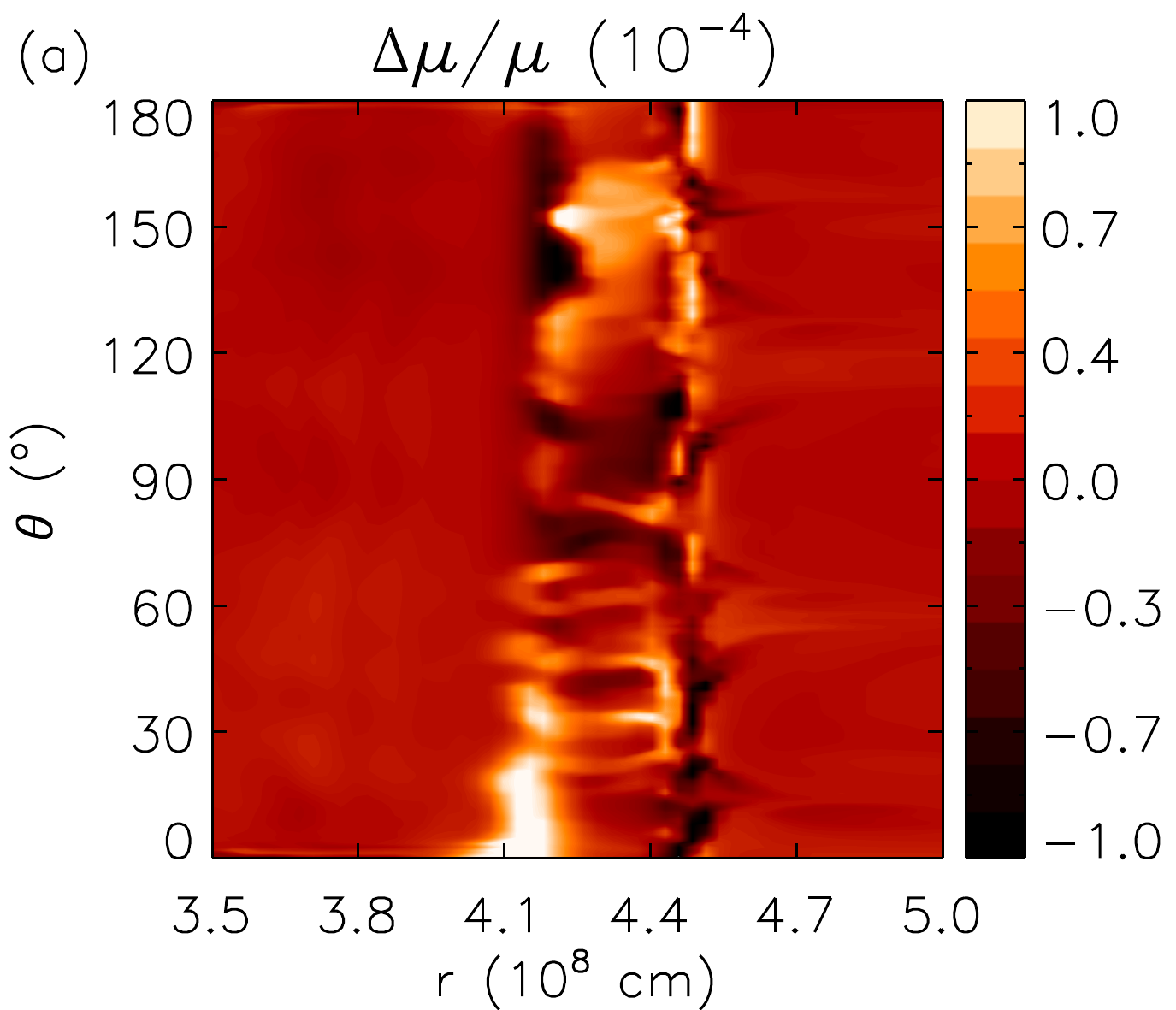}
\includegraphics[width=0.325\hsize]{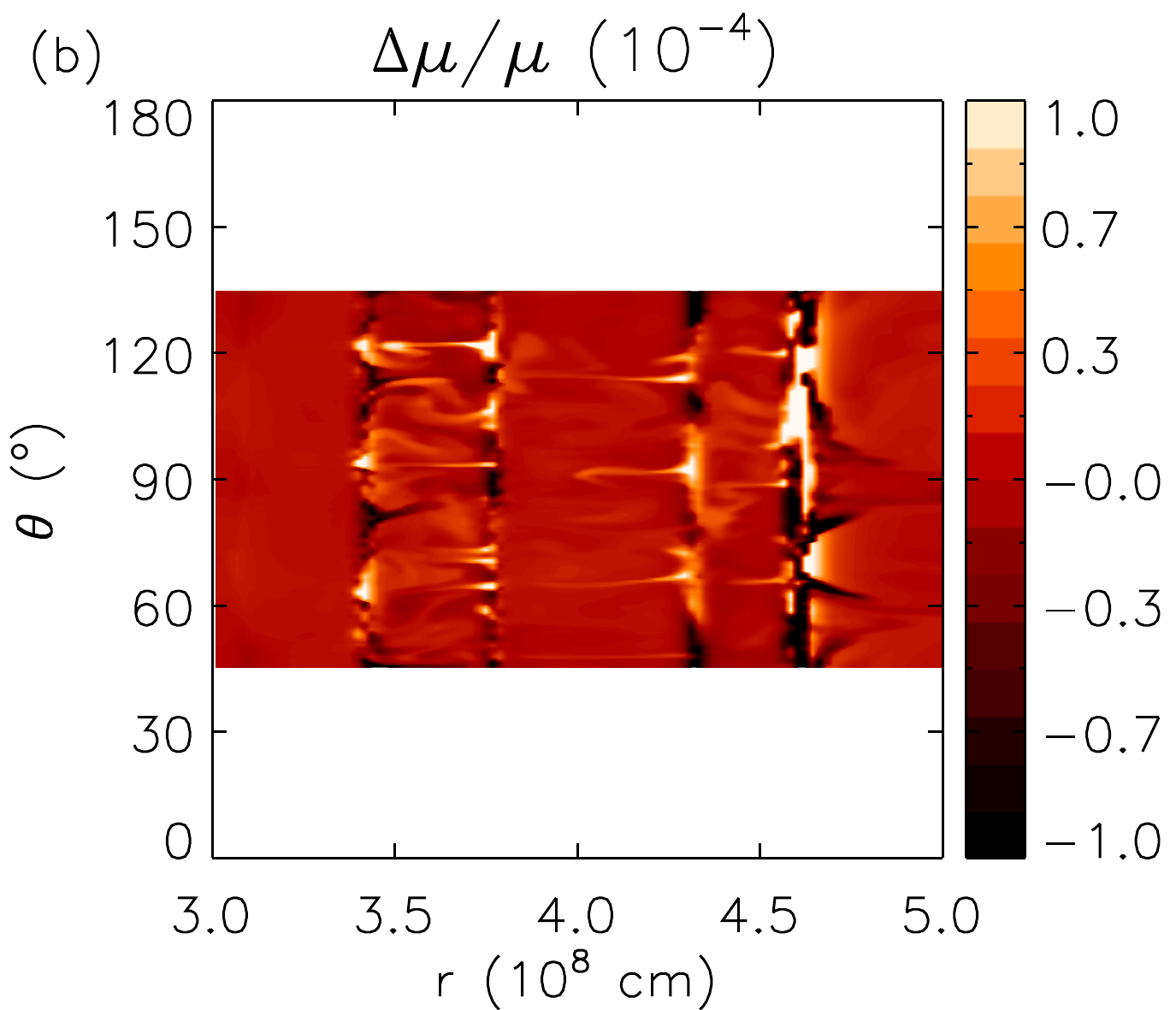}
\includegraphics[width=0.325\hsize]{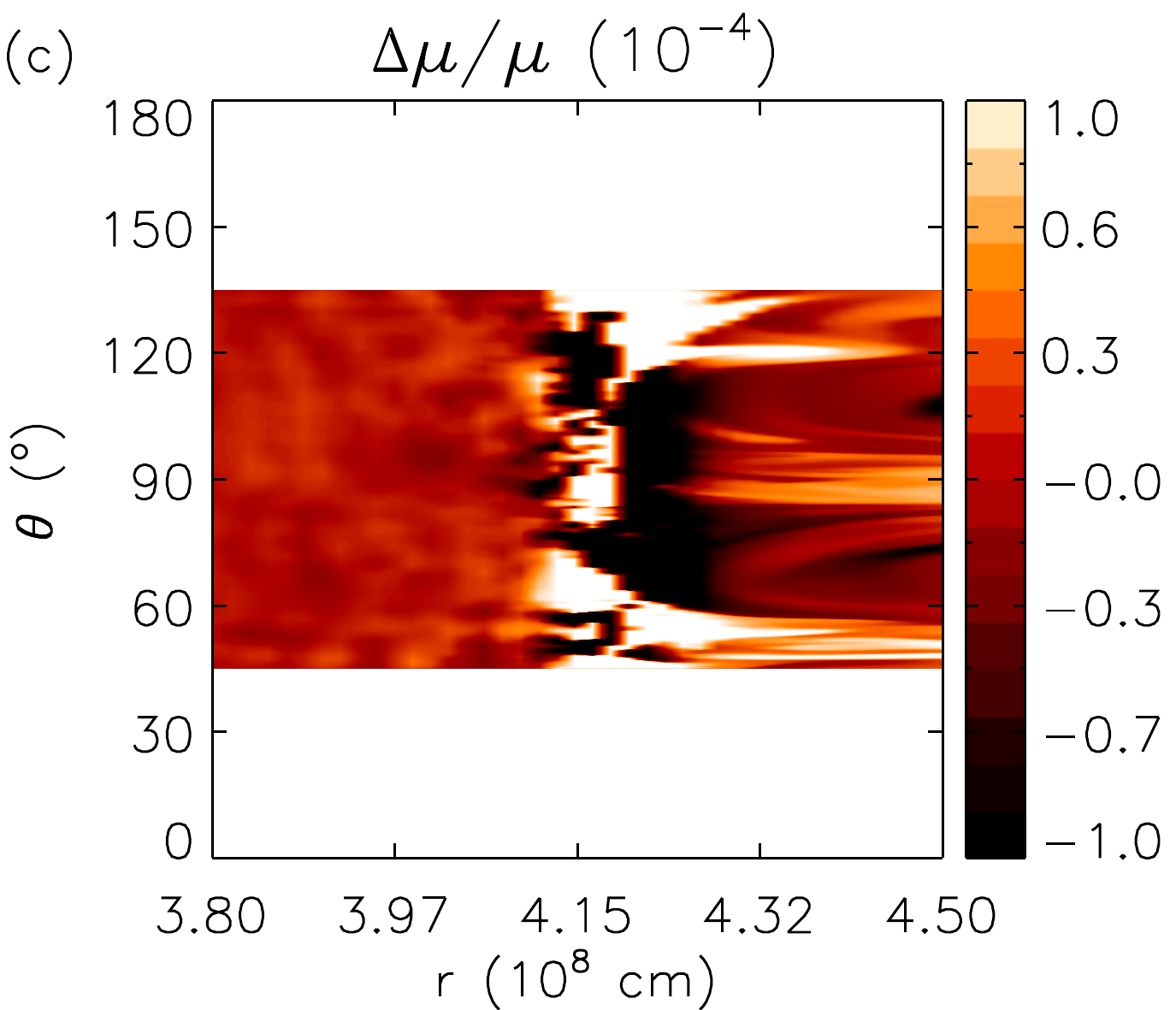}

\includegraphics[width=0.325\hsize]{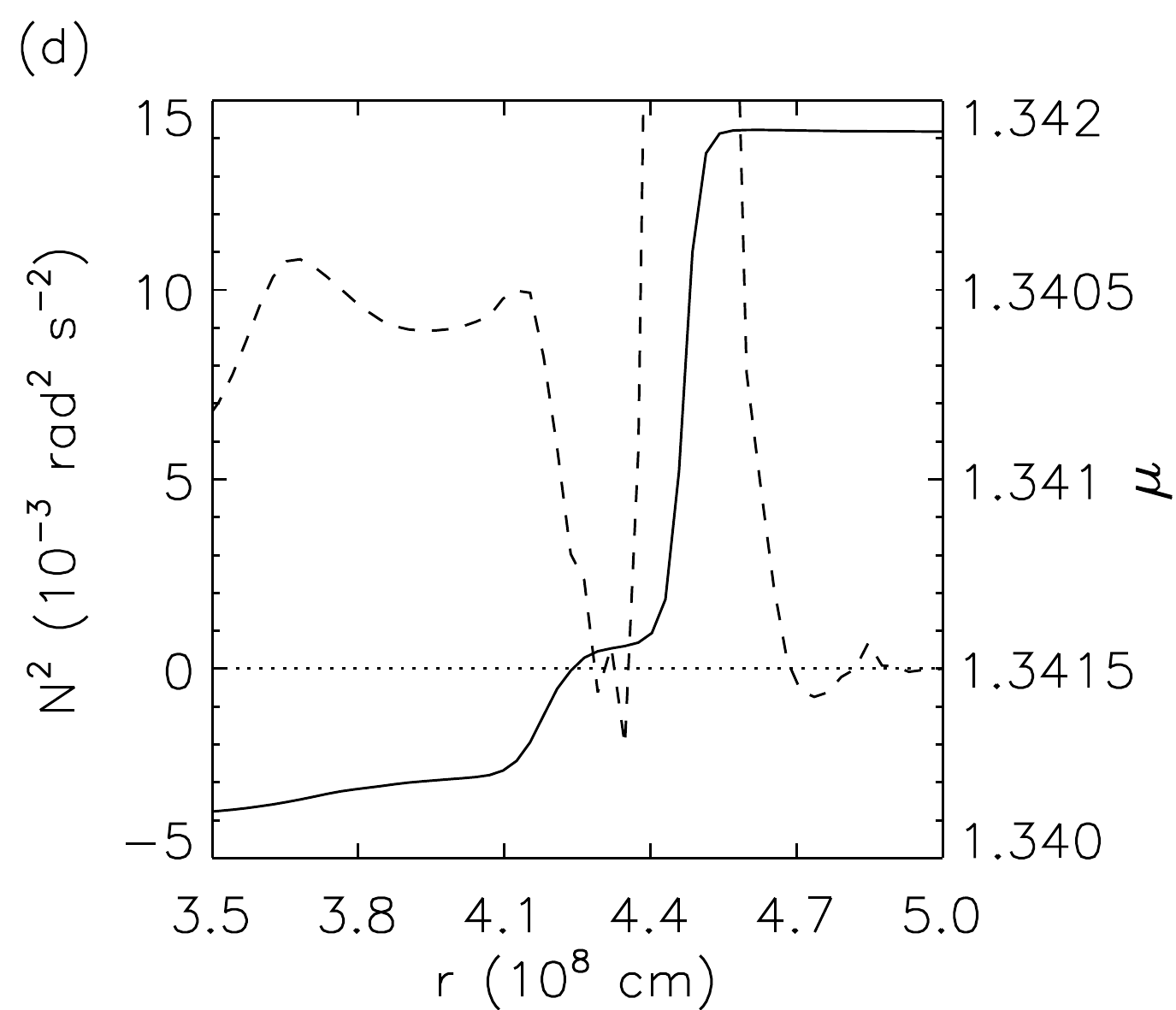}
\includegraphics[width=0.325\hsize]{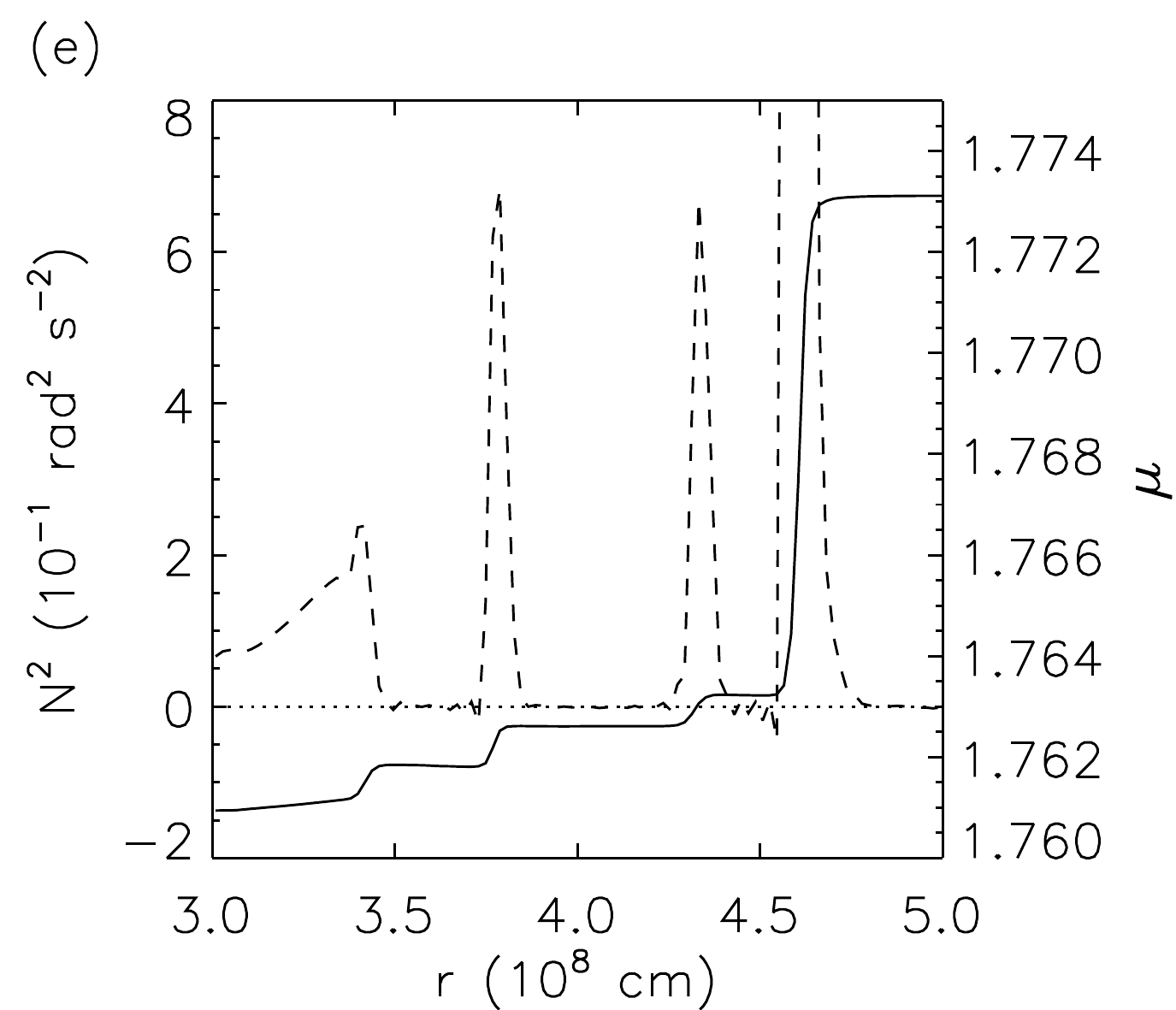}
\includegraphics[width=0.325\hsize]{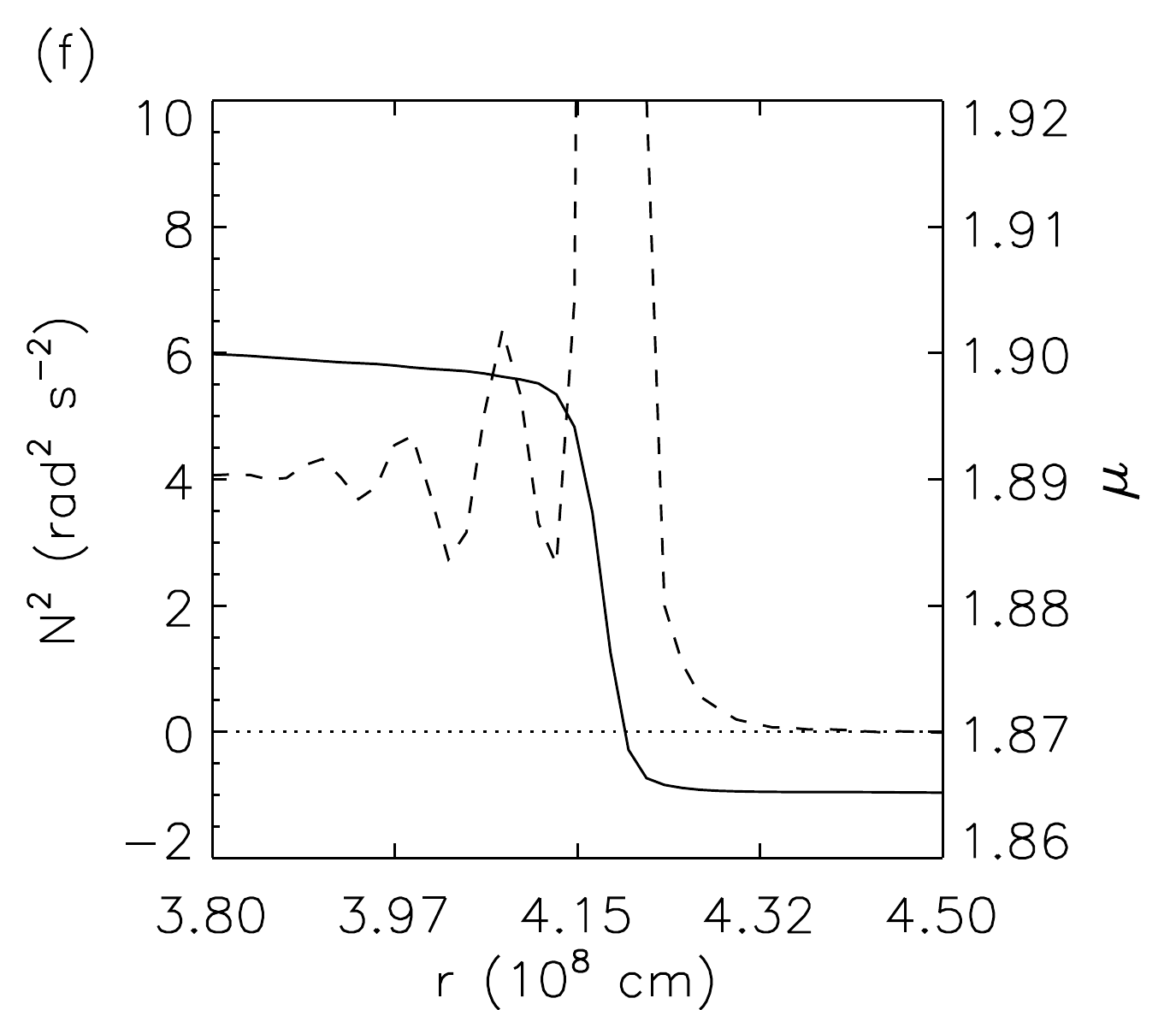}

\caption{Snapshots showing the relative angular fluctuations of the mean molecular weight $\Delta \mu/\mu = (\mu - \langle \mu \rangle_{\theta}) \,/\, \langle \mu \rangle_{\theta}$ at the base of the convection zone in the 2D model hefl.2d.3 and $t = 90076\,$s ~(a); cafl.2d and $t = 50890\,$s ~(b); oxfl.2d and $t = 4313\,$s~(c). Panels below (d,~e,~f) are related exactly to the ones above and shows a dashed line, which is the angular averaged distribution of the square of the Brunt-V\"ais\"al\"a frequency $N^2$, and the solid line is the angular averaged mean molecular weight $\mu$. The horizontal dotted line corresponds to $N^2 = 0$. $\langle \rangle_{\theta}$ denotes the angular average at a given radius.}
\label{fig.caflfing}
\end{figure*}

The layers just beneath the base of the convection zone show the presence of overdense sinking blob-like structures which are enriched with higher mean molecular weight $\mu$ material than the ambient matter into which they penetrate, thereby creating a compositional step as they redistribute the $\mu$.\footnote{Note that in Figs.\,\ref{fig.fingcarb} and \ref{fig.fingall}, which show the finger-like structures, the maps are displayed in spherical coordinates ($r$ and $\theta$ in cm and degree) using a rectangular projection. This makes the finger-like structures appear longer than they really are.} The beginning of new compositional step essentially coincides with the layer from which the mixing launches. Whereas it almost overlaps with the bottom of convection zone in early phases of the mixing (Fig.~\ref{fig.fingcarb};~panels $b$ and $c$), it stays clearly separated later (Fig.~\ref{fig.fingcarb};~panel $a$). The blobs are always cooler and denser than the ambient matter by roughly $0.4\%$ and $0.05\%$, respectively (Fig.\,\ref{fig.fingall}). 
Using the value of relative density fluctuations $\Delta \rho / \rho$ within the region where our mixing takes place we can derive an analytic estimate for the velocity of the dense blobs causing the finger-like structures, if we assume that the blobs arise from a dynamic instability \citep{CoxGiuli2008}. With $\Delta \rho / \rho \sim 5\times 10^{-4}$, inferred from our simulation (Fig.\,\ref{fig.fingall}\,a), we find
\begin{equation}
  v \sim \sqrt{g \Lambda \frac{\Delta \rho}{\rho}} \sim 7\times 10^5 \cms  
\label{eq:deltarho1}
\end{equation}
where $g \sim 10^8 \cmss$ is the gravitational acceleration at the base of the convection zone, and $\Lambda \sim 10^7 \cm$ the approximate length of the fingers. The above estimate agrees well with the results of an analysis of our simulation, where the gas inside the fingers sinks at a bit smaller velocities ranging from $\sim 1 \times 10^{5}$\cms up to $3.5 \times 10^{5}$\cms.  It confirms that the ``buoyancy work'' ($\sim v^2$) is consistent with the acceleration of gas seen in the simulation. While the initial background state is dynamically stable (according to Ledoux and Schwarzschild), these flow properties suggests that the mixing takes place on a dynamic timescale.  This apparent contradiction is addressed in \S\ref{sec:linear-stability}.  The development of the flow at $t\sim 9\times 10^4$ s is depicted in Figure~\ref{fig.caflfing}a,~d.

\par {\it Carbon Flash Convection. ---} A similar mixing process also appears in the carbon-flash model. A snapshot showing various quantities in the lower half of the carbon burning shell convection zone and the underlying stable layer is presented in Figure~\ref{fig.caflfing}b,~e and can be seen to develop in a similar manner to the He flash model.

\par In this model, however, the initial structure of the stable layers  is slightly different from that in the helium-flash model. When the model was mapped onto our hydrodynamic grid and allowed to adjust to hydrostatic equilibiurm, some narrow bands within the stable layers appeared which were unstable to convection.  This was due merely to the differences between the stellar evolution code and hydrodynamics grid as well as the fact that the stable layer was only marginally stable.   These narrow unstable regions quickly mixed, and stabilized, producing a slight stair stepping and shallow chemical discontinuities in the stable layer rather than a smooth profile as in the initial stellar model.  Besides this initial transient, however, the evolution between this model and the helium flash model proceed in the same manner.

\par {\it Oxygen Shell Burning. ---} In contrast to the helium- and carbon-flash convection cases, the stable layer residing beneath the oxygen shell burning zone does not undergo noticeable mixing over similar timescales.  This is an interesting observation because the overall structure of the oxygen burning shell is remarkably similar to the flash convection cases, except for the mean molecular weight gradient, which is positive in the oxygen burning case. This fact strongly suggests that the molecular weight gradient is responsible, or at least, connected to the mixing seen in the shell convection models. We also point out that the buoyancy frequency $N^2$ under convection shell is significantly higher than in the previous two cases indicating stronger dynamic stability (Figure~\ref{fig.caflfing}c,~f), which could simply suppress the mixing. 

\begin{figure} 
\includegraphics[width=0.99\hsize]{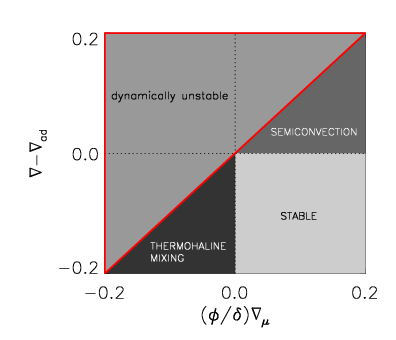}
\caption{Temperature ($\Delta\nabla = \nabla -\nabla_{ad}$) and composition gradient $(\phi/\delta)\nabla_\mu$ plane. The region above the diagonal is dynamically unstable according to the Ledoux criterion (Eq.\,\ref{eq:stability}).  The lower right quadrant is dynamically stable, while the "semiconvection" and "thermohaline" regions are dynamically stable, but unstable on thermal timescales.}
\label{fig:nabla}
\end{figure}

\subsection{Linear Stability of the Mixing Layer}
\label{sec:linear-stability}

\begin{figure*} 
\includegraphics[width=0.33\hsize]{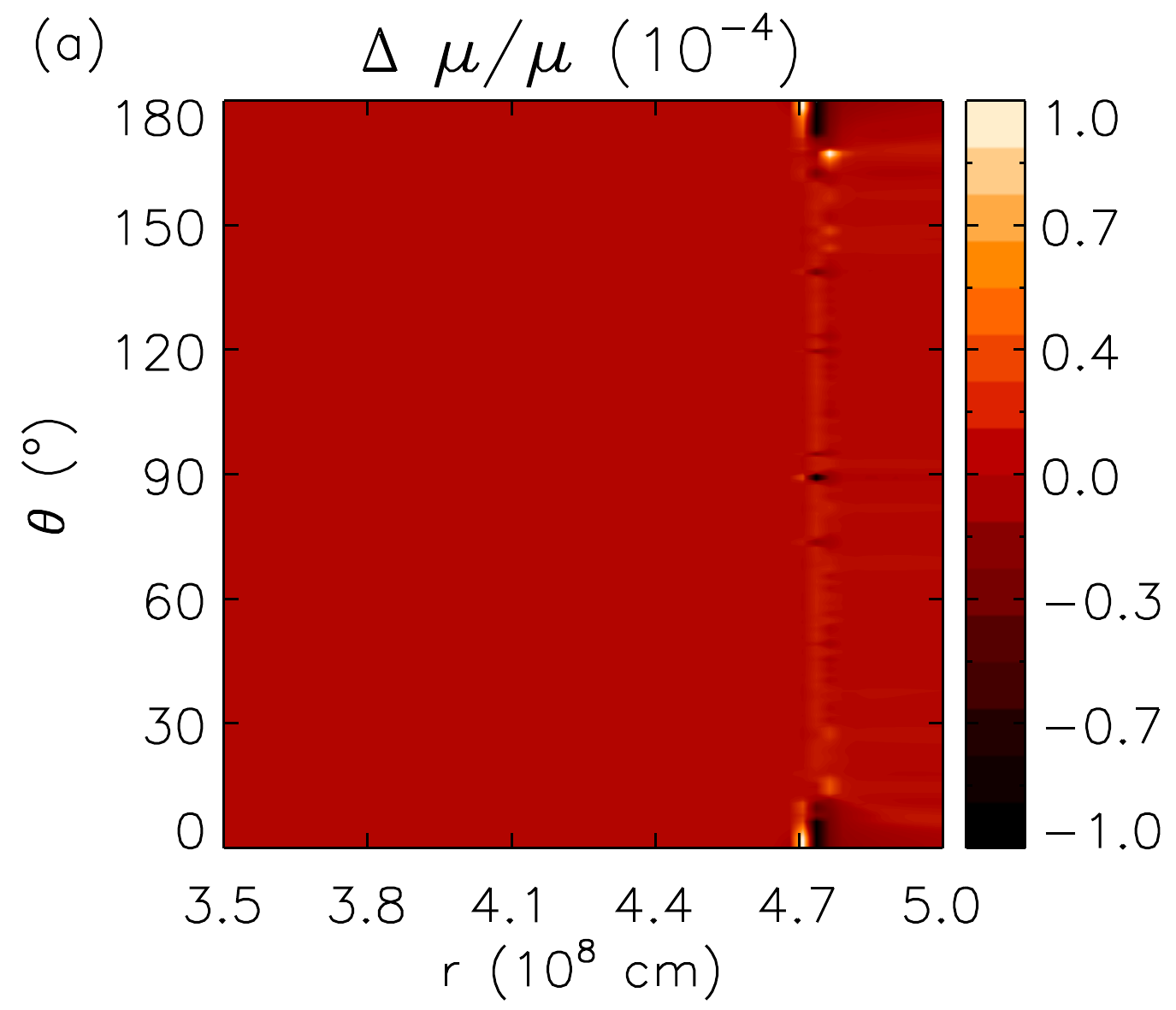}
\includegraphics[width=0.33\hsize]{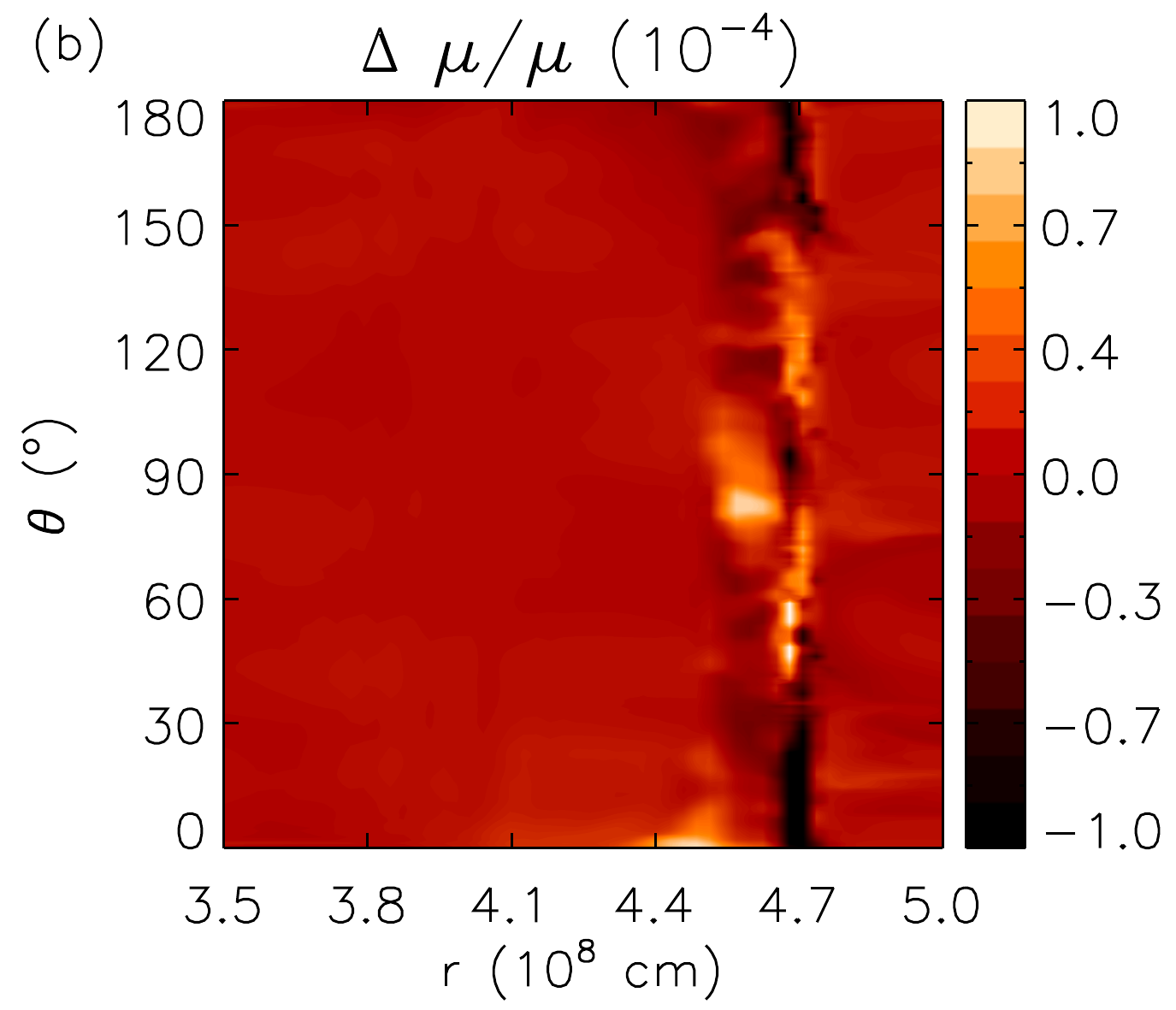}
\includegraphics[width=0.33\hsize]{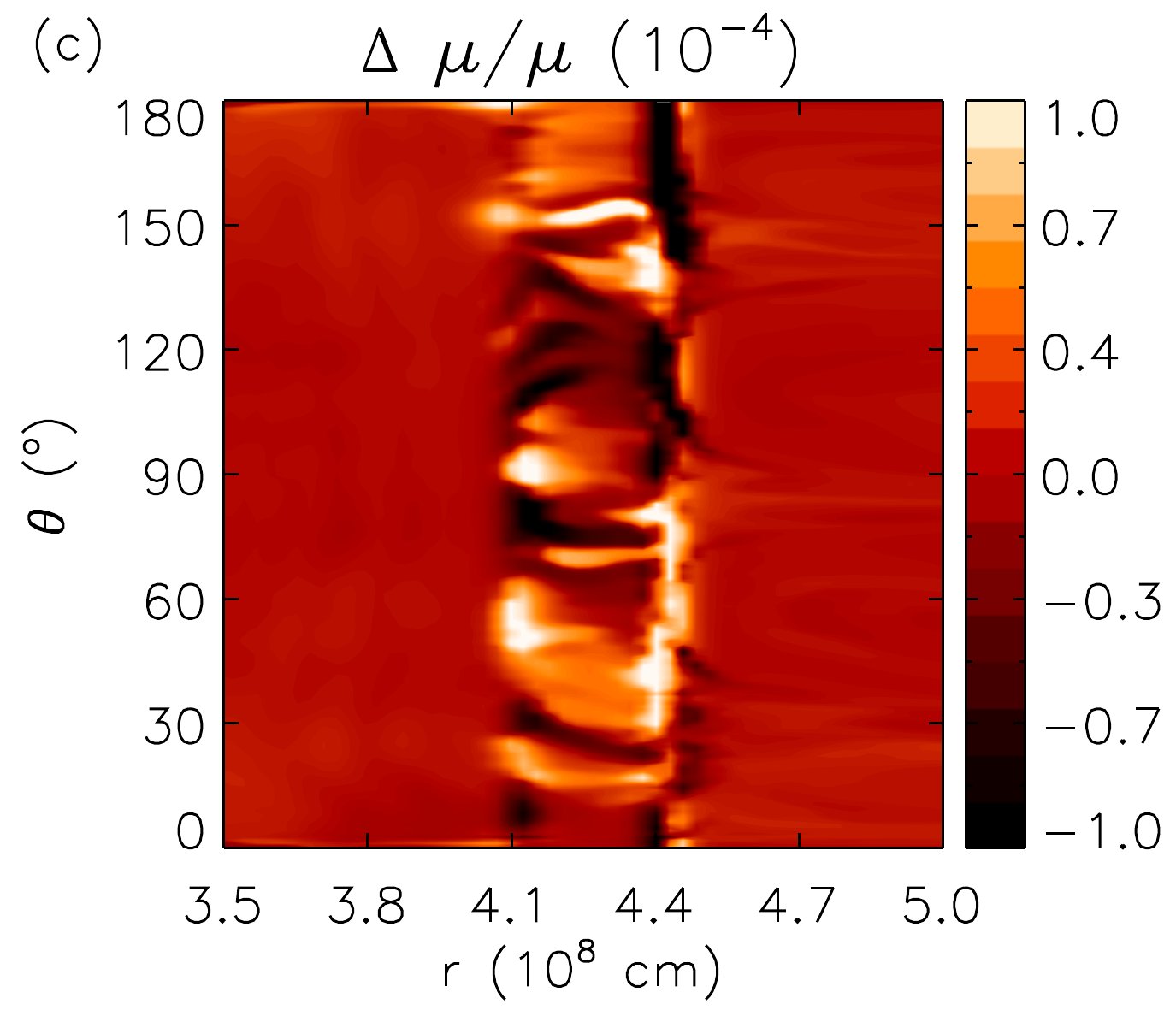} \\
\includegraphics[width=0.33\hsize]{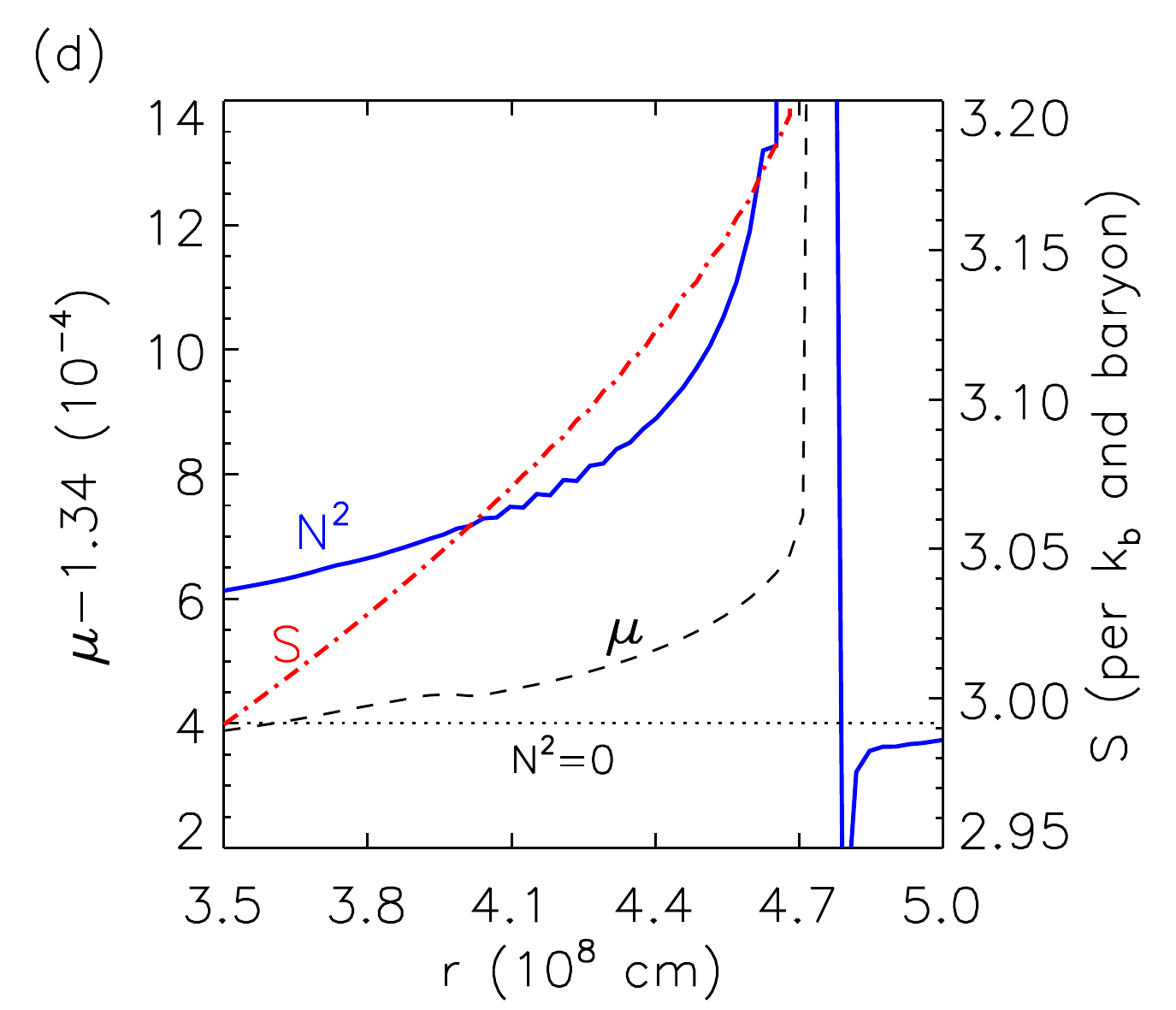}
\includegraphics[width=0.33\hsize]{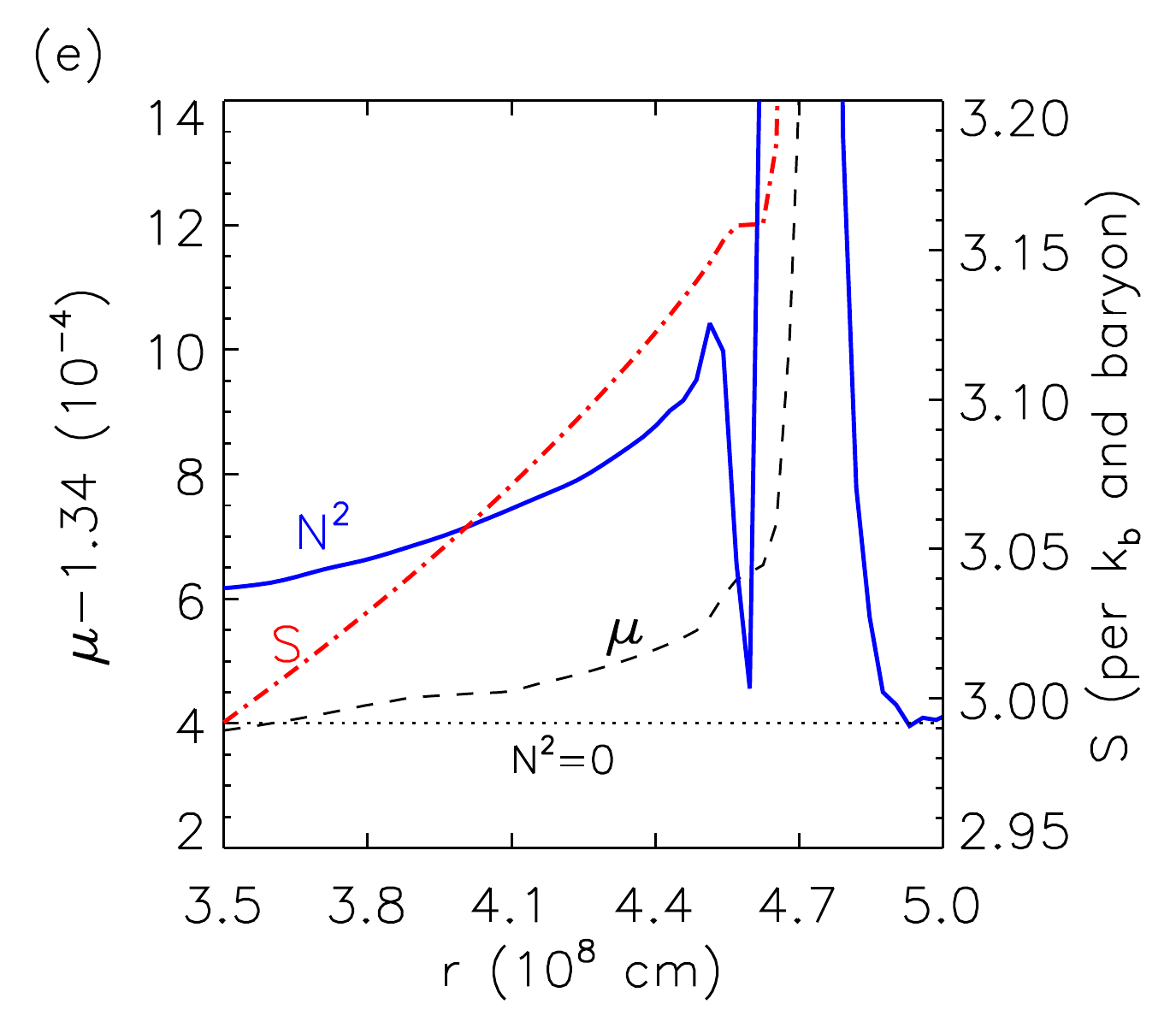}
\includegraphics[width=0.33\hsize]{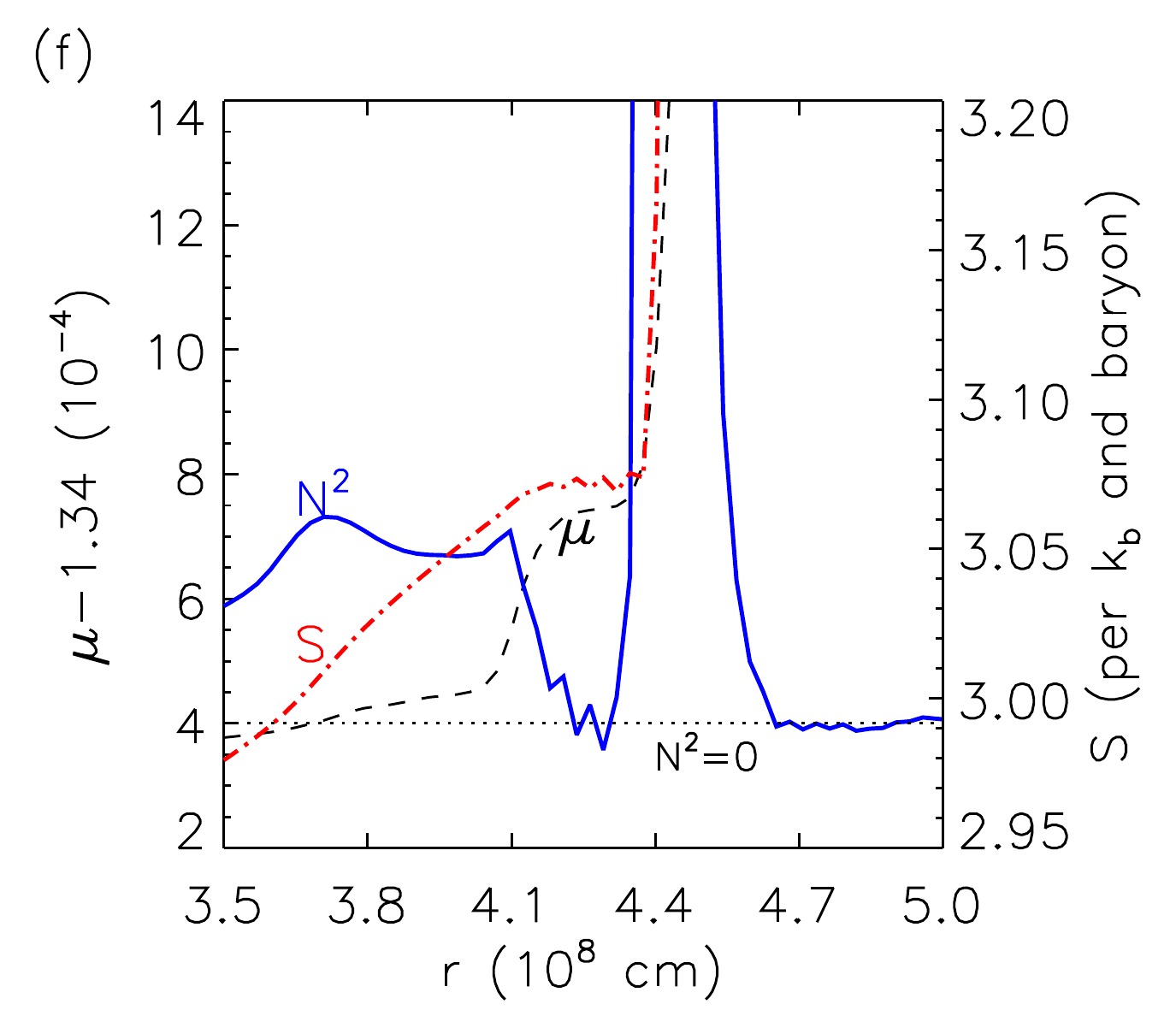}
\caption{A time sequence showing the evolution of relative difference between the local and the horizontally averaged value of mean molecular weight \ie $\Delta\mu / \mu$ (a,b,c), horizontally averaged square of the Brunt-V\"ais\"al\"a frequency $N^2$ (solid), entropy S (dash-dotted) and $\mu$ (dashed) in the stable layer residing beneath the convective shell in the He-flash model {\sf hefl.2d.3} at tree times t$_1 = 6 \times 10^2$~s (a,d), t$_2 = 10^4$~s (b,e) and t$_3 = 10^5$~s. (c,f). The horizontal dotted line marks zero level of $N^2$. }
\label{fig:mix-sequence-he}  
\end{figure*}

\par As a starting point to analyze the mixing we consider the linear stability of the regions in question, the standard approach to determine the mixing properties in 1D stellar evolution codes.  For adiabatic displacements, the condition for stability is often written in terms of the temperature and compositions gradients,  and reads

\begin{equation}
\label{eq:stability}
\Delta\nabla < (\varphi/\delta)\nabla_\mu
\end{equation}

\noindent where the stellar structure gradients are $\nabla = d\ln T/d\ln P$, $\nabla_{\mu} = d\ln\mu/d\ln P$, the thermodynamic derivatives are $\nabla_{ad} = (\partial\ln T/\partial\ln P)_{s,\mu}$, $\delta = -(\partial\ln\rho/\partial\ln T)_{P,\mu}$,  $\varphi = (\partial\ln\rho/\partial\ln\mu)_{P,s}$, and $\Delta\nabla = \nabla - \nabla_{ad}$.  This stability condition is known as the Ledoux criterion for convective stability \citep{KipWeigert1990} and it measures the degree to which the background is "superadiabatic".  In a homogeneous composition medium the stability condition reads $\Delta\nabla < 0$, which is known as the Schwarzschild criterion and is equivalent to the statement that positive entropy gradients ($ds/dr > 0$)  are stabilizing , negative entropy gradients are destabilizing, and zero entropy gradients result in neutral buoyancy upon displacement in the linear regime when pressure equilibrium is maintained \citep{LandauLifshitz1959}. The composition term in Eq.~\ref{eq:stability} accounts for the buoyancy due to mean molecular weight gradients in the medium. 

The stability condition (eq.[\ref{eq:stability}]) can be identified with the region in the "nabla-plane" below the line described by $\Delta\nabla = (\varphi/\delta)\nabla_{mu}$, as illustrated in Figure~\ref{fig:nabla}.  Above this line the fluid is subject to dynamical instability (e.g., thermal convection, Rayleigh-Taylor instabilities), while below the line the fluid is in a dynamically stable regime.  But what about non-adiabatic effects?  The regions in this plane labeled "semiconvection" and "thermohaline" are in fact unstable on thermal timescales due to secular instabilities\footnote{Secular instabilities are distinguished by growth timescales that are significantly longer than the dynamical timescale.}, and their linear growth properties have been analyzed for a sampling of astrophysical conditions \citep{Ulrich1972,Kippenhahn1980,Grossman1996,Charbonnel2007,Siess2009,Stancliffe2009}.  

\par \citet{Kippenhahn1980} first recognized that off-center helium and carbon ignition would indeed lead to thermohaline mixing below the burning shells but the long (thermal) timescale estimated for the growth of this instability compared to the rapid evolution of the shell flash led to a diminished interest in this context. In our shell flash simulations, the layers below the shell convection are indeed unstable according to the thermohaline condition ($\nabla<\nabla_{ad}$ and $\nabla_\mu < 0$) but {\em the mixing process operating in our simulations is in fact distinct from direct thermohaline mixing}, since it operates on a significantly shorter timescale than the thermal timescale.  Instead, we find that {\em the regions which undergo mixing are forced by waves excited within the region by the adjacent turbulent convection zone}.  Therefore, what we have captured in our simulations is a wave induced, turbulent mixing process operating in a stably stratified layer.

\subsection{Wave Driven Mixing}

\par As described in \S\ref{sec:general-features} and \S\ref{sec:linear-stability}, the regions immediately below the convective shells are initially (linearly) dynamically stable.  Furthermore, the absence of heat conduction in these simulations precludes these regions from undergoing thermally driven instabilities, such as semiconvection and thermohaline mixing. Therefore, the instabilities which develop in these regions are most naturally attributable to the jostling of the layers by turbulence in the adjacent convection zone.  The nature of the jostling is predominantly in the form of internal wave motions (i.e., gravity waves).

\par As the stable layer is jostled by the neighboring convection, mixing is instigated which redistributes the entropy and the composition. This is illustrated in the time sequence presented in Figure~\ref{fig:mix-sequence-he} for the helium shell flash model {hefl.2d.3}. 

\par There are several notable features in this figure.  First, the convection zone remains separated from the underlying stable layer by a spike in buoyancy frequency $N^2$ (around r$\sim 4.7\times 10^8$~cm in panels $a$ and $b$) arising from the entropy jump between the convection zone and the underlying layer. This entropy jump is due to the energy deposition by the nuclear burning in the shell. Note that the stabilizing layer is present despite the destabilizing $\mu-$profile. In other words, the entropy profile is stable enough to compensate for the destabilizing $\mu-$gradient effect. The spike in buoyancy also prevents overshoot mixing into the underlying stable layers.

\par Another notable feature is the reduction in stability just below the buoyancy frequency spike which separates the region from the convection zone as time progresses.  This reduction in stability (or decrease in N$^2$) is associated with a flattening of the entropy profile in the stable layer due to mixing induced in the region. The mixing that ensues is also associated with a modification of the composition profile, and by  $t\sim10^4$ seconds a new step in the $\mu$-profile has already formed near $r\sim 4.5\times10^8$ cm. 

\par Finally, it is worth pointing out that in addition to the mixing inside the stable layer there is an overall deepening of the convection zone due to turbulent entrainment.  By $t\sim 9.5\times10^4$ seconds the convective boundary has migrated inward a distance of $\Delta r\sim 2\times10^7$cm, from $r\sim 4.8\times10^8$ cm to $r\sim 4.6\times10^8$ cm.

\par In the stable layers, the mixing is strongest in regions where the gradients are largest, as expected for a diffusive mixing process. Since compositional mixing at the molecular level is not included in these simulations, nor is heat conduction, this mixing results as a consequence of numerical diffusion. We stress here that the numerical diffusion can have a realistic counterpart in stars, but caution should be paid to the quantitative rate of mixing presented in this paper. We will discuss the connection between the simulated results (and numerical mixing) and the situation in real stars below in terms of wave enhanced diffusion as studied by, e.g., \citet{press1981}.

\subsection{Layering and Interface Migration}

\par
\begin{figure} 
\includegraphics[width=0.99\hsize]{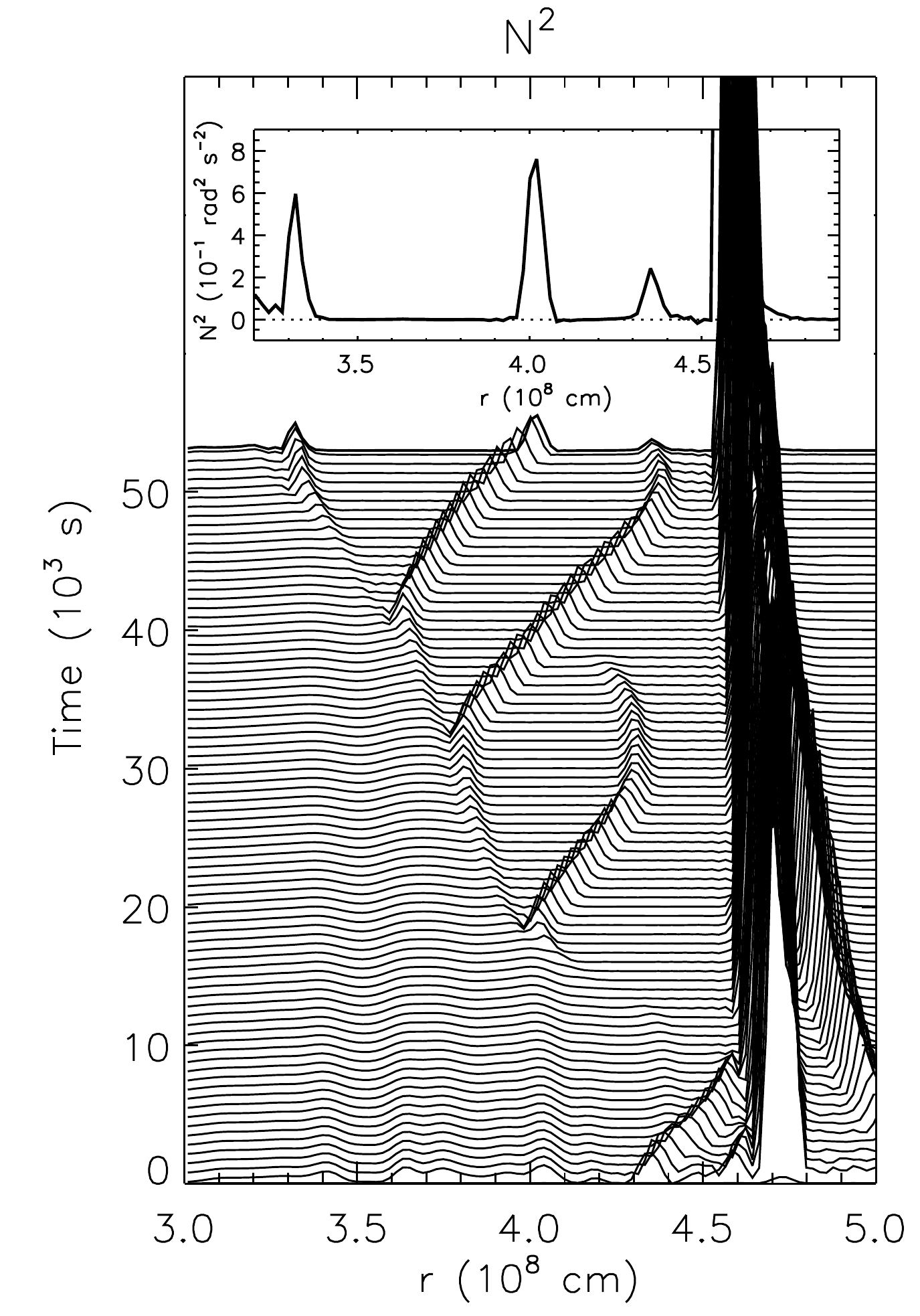}
\caption{Profiles of the buoyancy frequency N$^2$ as a function of radial coordinate for different values of the time. The vertical axis is time and the profiles are shown spaced at time intervals of $\sim$ 700 seconds. The inset figure at the top is showing the scale for N$^2$ corresponding to the last profile at time $\sim$ 53000~s.}
\label{fig.n2}
\end{figure}

 A striking phenomena seen in both the He- and C-flash convection models is the formation of homogeneous layers in the mixing region separated by steep gradient interfaces.  These are apparent in both  Figure~\ref{fig.caflfing} and Figure~\ref{fig:mix-sequence-he}.  Once these layers have formed, a subsequent phase of evolution takes place which involves the migration of the interfaces separating these layers.  This is illustrated in the series of snapshots compiled in Figure~\ref{fig.n2}, which shows the time evolution of the buoyancy frequency $N^2$ in the mixing region for the carbon flash model. These $N^2$ profiles and their peaks are seen to evolve by a process of corresponding interface migration driven by turbulent entrainment.  This entrainment is powered by the turbulent kinetic energy deposited in the region by trapped gravity waves which are excited by the overlying convection zone, rather than thermally driven turbulence as in the case of a convection zone.

\par The formation of steps and interface migration is a well known phenomena in a variety of fluid dynamic systems, including those studied in geophysical systems, in laboratories, and in simulation.  A very close analog to the type of interface migration that we see in our simulations is the situation modeled by \citet[][see their Figure 6 and 13, which show striking similarities to our Figure~\ref{fig.n2}]{Balmforth1998}. In their work, the evolution of buoyancy due to turbulence in a stably stratified layer is modeled in an attempt to understand the origin of layer formation and the dynamics of interface migration. This  situation is a very close analog to the mixing process taking place in our simulations, except that the source of turbulence induced in the mixing region in our calculations is due to wave motions excited in the region.  

\par Although the quantitative rate of mixing in these regions remains to be identified through higher resolution calculations and analysis, the formation of steps and their migration through the stable layer provide important clues about how one might treat this mixing in 1D stellar evolution codes \citep[e.g.][]{Balmforth1998,zilitenkevich2007}.   

\section{Discussion and Conclusions}
\label{sect:sum}

\par We have identified a mixing process operating in our simulations of off-center helium- and carbon-flash convection. The mixing results in the redistribution of composition and entropy within the stable layers that reside beneath the shell convection zones. Associated with the mixing process is the formation of "fingers" of high mean molecular weight material which traverse the mixing region.  These "fingers" are essentially tracer of material trasport taking place across the stable layer (e.g., Figure~\ref{fig.fingcarb}).

\par We find that this mixing takes place against a background state which is initially {\em dynamically stable} to linear perturbations.  And while the mixing region is indeed unstable to thermohaline mixing on a secular timescale, the mixing process that we observe operates on a {\em dynamical timescale}, thus excluding thermohaline mixing as the root cause.  This is reinforced by the fact that heat conduction was not included in these calculations, primarily because its effects are too slow to operate over the time frames simulated. The thermal timescale for thermohaline mixing \citep{Kippenhahn1980} estimated for our core helium flash model is of the order of $10^5$~years. It is neither expected that the timescale will drastically change (decrease) with improved stellar models nor during the flash as the thermal structures are in general very similar \citep{SweigertGross1978}.

\par Our search for a counterpart to the mixing observed in our simulations among terrestrial experiments with qualitatively similar conditions led us to phenomena like viscous or density fingering driven by a Rayleigh-Taylor instability \citep{AnneDeWit2004,AnneDeWit2008} and buoyancy-driven instabilities resulting from differences in mass diffusion coefficients of chemically-reacting species sitting on top of each other \citep{Almarcha2010}. Properties of these phenomena do not fit the mixing in our simulations.

\par As an alternative explanation we describe a scenario in which mixing is instigated by the finite amplitude jostling motions of the stable layer by the turbulence present in the adjacent convective shell. The perturbations induced in the stable layer by the adjacent convection are primarily in the form of waves (i.e., g-modes).  The jostling of the stable layer results in a diffusive mixing process therein which leads to a redistribution of the entropy and composition on a dynamical timescales. This scenario is qualitatively similar to the enhanced mixing discussed by \citet{press1981} and \cite{pressrybicki1981} who found that thermal and mass transport is not only a function of temperature and composition gradients, but also a function of the fluid shearing motions setup by the presence of a passing wave field.

\par Indeed, the layer from which the mixing begins coincides with layers of strongest shear (Fig.\,\ref{fig.fingall}~c). But the shear and the finite amplitude perturbations alone do not drive mixing and have to be preceded by the formation of sufficiently steep compositional gradients. In the case of our core helium and carbon flash models, such gradients are inherent in our initial stellar models at the bottom of convection shells due to off-center ignition. 

\par The biggest source of uncertainty inherent in this work is the quantitative rate of mixing.  This uncertainty stems from the fact that the simulated mixing results from enhanced numerical diffusion, rather than from a resolved velocity field and molecular diffusion processes.  Higher resolution simulations as well as analytic modeling should be pursued, perhaps in restricted models, in order to quantify the mixing rate due to this process.  This is necessary for including it into a stellar evolution code.  

\par One of the pressing goals for stellar evolution is the construction of algorithms for 1D codes that capture mixing processes such as that described in this paper, and which extend mixing due to hydrodynamic processes beyond the boundaries of convective regions and into stably stratified layers.  The need to treat turbulence and mixing in stable layers has already been recognized in the geophysical community \citep[e.g.,][]{Fernando1991,zilitenkevich2007}, and this work can serve to inform a related effort for stellar evolution.

%
\begin{acknowledgements}
The simulations were performed at the computer center of the Max-Planck-Society in Garching (RZG). Miroslav\,Moc\'ak acknowledges financial support from the Communaut\'e francaise de Belgique - Actions de Recherche Concert\'ees, and from the Institut d'Astronomie et d'Astrophysique at the Universit\'e Libre de Bruxelles (ULB). The authors want to thank to Christophe Almarcha and Anne De Wit for enlightning discussions on chemo-hydrodynamics. We express our gratitude to Rob Izzard for valuable discussions, and for helpful comments on the manuscript. 
\end{acknowledgements}



\begin{thebibliography}{28}
\expandafter\ifx\csname natexlab\endcsname\relax\def\natexlab#1{#1}\fi

\bibitem[{{Almarcha} {et~al.}(2010){Almarcha}, {Trevelyan}, {Riolfo}, {Zalts}, {Hasi}, {D'Onofrio}, \& {De Wit}}]{Almarcha2010} {Almarcha}, C., {Trevelyan}, P.~M.~J., {Riolfo}, L.~A., {Zalts}, A., {Hasi}, C.~E., {D'Onofrio}, A.~D., \& {De Wit}, A. 2010, The Journal of Physical Chemistry Letters, 1, 752

\bibitem[{{Arnett} {et~al.}(2009){Arnett}, {Meakin}, \& {Young}}]{amy2009}
{Arnett}, D., {Meakin}, C., \& {Young}, P.~A. 2009, \apj, 690, 1715

\bibitem[{{Arnett} {et~al.}(2010){Arnett}, {Meakin}, \& {Young}}]{amy2010}
{Arnett}, D., {Meakin}, C., \& {Young}, P.~A. 2010, \apj, 710, 1619


\bibitem[{{Balmforth} {et~al.}(1998){Balmforth}, {Smith}, \&
  {Young}}]{Balmforth1998}
{Balmforth}, N.~J., {Smith}, S. G.~L., \& {Young}, W.~R. 1998, J.Fluid Mech.,
  335, 329

\bibitem[{{Charbonnel} \& {Zahn}(2007)}]{Charbonnel2007}
{Charbonnel}, C., \& {Zahn}, J. 2007, \aap, 467, L15

\bibitem[{{Colella} \& {Glaz}(1984)}]{ColellaGlaz1984}
{Colella}, P., \& {Glaz}, H.~H. 1984, J.Comp.Phys., 59, 264

\bibitem[{{Colella} \& {Woodward}(1984)}]{ColellaWoodward1984}
{Colella}, P., \& {Woodward}, P.~R. 1984, J.Comp.Phys., 54, 174

\bibitem[{{Dearborn} {et~al.}(2006){Dearborn}, {Lattanzio}, \& {Eggleton}}]{dle2006}
{Dearborn}, D.~S.,~P., {Lattanzio}, J.~C., \& {Eggleton}, P.~P. 2006, \apj, 639, 405

\bibitem[{{De Wit}(2004)}]{AnneDeWit2004}
{De Wit}, A. 2004, These d'Agregation de l'Enseignement Superieur
  (Habilitation), 1, 0

\bibitem[{{De Wit}(2008)}]{AnneDeWit2008}
---. 2008, Chimie Nouvelle, 99, 1

\bibitem[{{Fernando}(1991)}]{Fernando1991}
{Fernando}, H. 1991, Ann.Rev.Fluid Mech., 23, 455

\bibitem[{{Grossman} \& {Taam}(1996)}]{Grossman1996}
{Grossman}, S.~A., \& {Taam}, R.~E. 1996, \mnras, 283, 1165

\bibitem[{{Herwig} {et~al.}(2006){Herwig}, {Freytag}, {Hueckstaedt}, \&
  {Timmes}}]{Herwig2006}
{Herwig}, F., {Freytag}, B., {Hueckstaedt}, R.~M., \& {Timmes}, F.~X. 2006,
  \apj, 642, 1057

\bibitem[{{Herwig} {et~al.}(2011){Herwig}, {Pignatari}, {Woodward}, {Porter}, {Rockefeler}, {Fryer}, {Bennett}, {Hirschi}}]{Herwig2011}
{Herwig}, F., {Pignatari}, M., {Woodward}, P.~R., {Porter}, D.~H., {Rockefeller}, G.,  {Fryer}, C.~L., {Bennett}, M., {Hirschi}, R. 2011, \apj, 727, 88


\bibitem[{{Kippenhahn} {et~al.}(1980){Kippenhahn}, {Ruschenplatt}, \&
  {Thomas}}]{Kippenhahn1980}
{Kippenhahn}, R., {Ruschenplatt}, G., \& {Thomas}, H. 1980, \aap, 91, 175

\bibitem[{{Kippenhahn} \& {Weigert}(1990)}]{KipWeigert1990}
{Kippenhahn}, R., \& {Weigert}, A. 1990, {Stellar Structure and Evolution}
  (Stellar Structure and Evolution, XVI, 468 pp.~192 figs..~ Springer-Verlag
  Berlin Heidelberg New York.~Also Astronomy and Astrophysics Library)

\bibitem[{{Landau} \& {Lifshitz}(1959)}]{LandauLifshitz1959}
{Landau}, L.~D., \& {Lifshitz}, E.~M. 1959, {Fluid mechanics}, ed. {Landau,
  L.~D.~\& Lifshitz, E.~M.}

\bibitem[{{Meakin} \& {Arnett}(2007)}]{MeakinArnett2007}
{Meakin}, C.~A., \& {Arnett}, D. 2007, \apj, 667, 448

\bibitem[{{Moc{\'a}k} {et~al.}(2008){Moc{\'a}k}, {M{\"u}ller}, {Weiss}, \&
  {Kifonidis}}]{Mocak2008}
{Moc{\'a}k}, M., {M{\"u}ller}, E., {Weiss}, A., \& {Kifonidis}, K. 2008, \aap,
  490, 265

\bibitem[{{Moc{\'a}k} {et~al.}(2009){Moc{\'a}k}, {M{\"u}ller}, {Weiss}, \&
  {Kifonidis}}]{Mocak2009}
---. 2009, \aap, 501, 659

\bibitem[{{Moc{\'a}k} {et~al.}(2010){Moc{\'a}k}, {Campbell}, {M{\"u}ller}, \&
  {Kifonidis}}]{Mocak2010}
{Moc{\'a}k}, M., {Campbell}, S.~W., {M{\"u}ller}, E., \& {Kifonidis}, K. 2010, \aap, 520, 114

\bibitem[{{Plewa} \& {M{\"u}ller}(1999)}]{PlewaMueller1999}
{Plewa}, T., \& {M{\"u}ller}, E. 1999, \aap, 342, 179

\bibitem[{{Press}(1981)}]{press1981}
{Press}, W.~H. 1981, \apj, 245, 286

\bibitem[{{Press} \& {Rybicki}(1981)}]{pressrybicki1981}
{Press}, W.~H., \& {Rybicki}, G.~B. 1981, \apj, 248, 751

\bibitem[{{Siess}(2006)}]{Siess2006}
{Siess}, L. 2006, \aap, 448, 717

\bibitem[{{Siess}(2009)}]{Siess2009}
---. 2009, \aap, 497, 463

\bibitem[{{Stancliffe} {et~al.}(2009){Stancliffe}, {Church}, {Angelou}, \&
  {Lattanzio}}]{Stancliffe2009}
{Stancliffe}, R.~J., {Church}, R.~P., {Angelou}, G.~C., \& {Lattanzio}, J.~C.
  2009, \mnras, 396, 2313

\bibitem[{{Sweigart} \& {Gross}(1978)}]{SweigertGross1978}
{Sweigart}, A.~V., \& {Gross}, P.~G. 1978, \apjs, 36, 405

\bibitem[{{Ulrich}(1972)}]{Ulrich1972}
{Ulrich}, R.~K. 1972, \apj, 172, 165

\bibitem[{Weiss {et~al.}(2004)Weiss, Hillebrandt, Thomas, \&
  Ritter}]{CoxGiuli2008}
Weiss, A., Hillebrandt, W., Thomas, H.-C., \& Ritter. 2004, Cox and Giuli's
  Principles of Stellar Structure ({Gardners Books})

\bibitem[{{Weiss} \& {Schlattl}(2000)}]{WeissSchlattl2000}
{Weiss}, A., \& {Schlattl}, H. 2000, \aaps, 144, 487

\bibitem[{{Weiss} \& {Schlattl}(2007)}]{WeissSchlattl2007}
---. 2007, \apss, 341

\bibitem[{{Zilitinkevich} {et~al.}(2007){Zilitinkevich}, {Elperin}, {Kleeorin},
  \& {Rogachevskii}}]{zilitenkevich2007}
{Zilitinkevich}, S.~S., {Elperin}, T., {Kleeorin}, N., \& {Rogachevskii}, I.
  2007, Boundary-Layer Meteorology, 125, 167

\end{thebibliography}
\end{document}